\documentclass[11pt]{article}
\usepackage[left=3cm, right=3cm, top=3cm, bottom=3cm]{geometry}
\usepackage{amsmath,amsfonts,amsthm,amssymb}
\usepackage{graphicx}
\usepackage{tikz}
\usepackage{bm}
\usepackage{array}
\usepackage{multirow}
\usepackage{hhline}
\usepackage{empheq}
\usetikzlibrary{arrows.meta}
\usepackage{mathrsfs}
\usepackage{authblk}
\usepackage{comment}

\def\angmom{\bm{c}}
\def\bq{\bm{q}}
\def\qhat{\hat{\bm{q}}}

\def\rhohat{{\hat{\bm{\rho}}}}
\def\br{\bm{r}}

\def\pobs{\bm{p}_{\rm obs}}
\def\rhovec{{\bm{\rho}}}

\def\erre{{\bm{r}}}
\def\erredot{\dot{\erre}}

\def\angmom{{\bm{c}}}

\def\etre{{\bm e}_3}

\def\adj{\text{adj}}

\def\obs{{\rm obs}}
\def\dbq{\dot{\bm{q}}}
\def\bqperp{\bm{q}^\perp}
\def\dbqperp{\frac{d}{dt}{\bm{q}}^\perp}

\def\obs{{\rm obs}}

\newtheorem{remark}{\bf Remark}

\title{\bf Generalization of a method by Mossotti for initial orbit determination}

\author[1]{Giovanni F. Gronchi}
\author[1]{Giulio Ba\`u}
\author[1]{\'Oscar Rodr\'iguez}
\author[2]{Robert Jedicke}
\author[3]{Joachim Moeyens}
\affil[1]{Dipartimento di Matematica, Universit\`a di Pisa, Italy}
\affil[2]{Institute for Astronomy, University of Hawai`i, USA}
\affil[3]{Vera C. Rubin Observatory, University of Washington, USA}

\begin{document}
\maketitle

\begin{abstract}
Here we revisit an initial orbit determination method introduced by O. F. Mossotti employing four geocentric sky-plane observations and a linear equation to compute the angular momentum of the observed body.  
We then extend the method to topocentric observations,  yielding a quadratic equation for the angular momentum. 
The performance of the two versions are compared through numerical tests with synthetic asteroid data using different time intervals between consecutive observations and different astrometric errors.
We also show a comparison test with Gauss's method using simulated observations with the expected cadence of the VRO-LSST telescope.
\end{abstract}

\section{Introduction}

In 1816 Ottaviano F. Mossotti introduced a method for initial orbit determination of a solar system body employing four optical observations, e.g. the values of right ascension and declination. 
Assuming {\em geocentric} observations, Mossotti's method allows to write linear equations for the computation of the orbital angular momentum \cite{mossotti}. 
Then the orbit can be reconstructed, e.g. by Gibbs' method \cite{Herrick}. 
This procedure has the advantage to avoid the computation of the roots of the eight degree polynomial appearing in the classical methods by Laplace \cite{laplace}, Lagrange \cite{lagrange}, and Gauss \cite{gauss1809}, which need only three observations but can give rise to multiple solutions. 
A review of the methods by Laplace and Gauss together with a geometric interpretation of the occurrence of multiple solutions can be found in \cite{gronchi2009}, \cite{mg2010}.

Mossotti's work was appreciated by Gauss himself, see \cite{gausswerkeVI}.  
This method has been reviewed in \cite{cellpinz}, where the authors state that the computation of the solution can be seriously affected by the observational errors due to the terms that are neglected in the employed approximation.

In this work we recall Mossotti's original method and show that it is possible to define a {\em topocentric} version which leads to a quadratic equation for the angular momentum.
This generalization of the method turns out to be suitable for orbit determination of Earth satellites too. 
We investigate the performance of the methods and their sensitivity to observational errors by some numerical tests with simulated data: we compare the original geocentric method with this topocentric version using different time intervals between the observations and different astrometric errors. 
We also show a comparison test with Gauss's method using simulated observations with the expected cadence of the VRO-LSST telescope.

\section{The original method}
\label{s:mossotti_geo}

Mossotti's method \cite{mossotti} leads to a set of linear equations for the components of $\angmom_\oplus-\angmom$, where $\angmom_\oplus$ and $\angmom$ are the angular momenta of the Earth and a solar system body, respectively. 
The observations are supposed to be made from the center of the Earth. 
We assume that the observed body is an asteroid, moving along an elliptic Keplerian trajectory with the Sun as the center of force. 
We also assume that the total observational arc is covered in a much shorter time than the orbital period. 
This method is also suitable to be used with hyperbolic or parabolic orbits.

Here we illustrate all the formulae which are necessary for a numerical implementation following Mossotti's paper steps \cite{mossotti}. 
However, in the early XIXth century Linear Algebra had not been developed yet, and several formulae in \cite{mossotti} can be written and derived in a shorter way.  
The original formulae can be recovered using Table~\ref{compare} in Appendix~\ref{append}.

\subsection{Units and preliminary definitions}

In order to simplify the notation we use the rescaled time $\theta$, defined by
\begin{equation*}
    \theta = t\sqrt{g\left(1+\frac{m_\oplus}{m_\odot}\right)},
\end{equation*}
where $m_\oplus$ and $m_\odot$ are the masses of the Earth and the Sun, and $g=Gm_\odot$, where $G$ is Newton's gravitational constant.
We also use $\kappa$ to denote Gauss's constant $\sqrt{g}$.
In the following we assume that
\begin{equation*}
    \theta \approx \kappa t,
\end{equation*}
neglecting the constant ${m_\oplus}/{m_\odot}$.

Take three of the four observations, at epochs $t_1<t_2<t_3$, and set
\begin{equation*}
    \theta_{ij} = \kappa (t_j - t_i),
\end{equation*}
\begin{equation*}
    \bm{\theta} = (\theta_{23},\,\theta_{31},\,\theta_{12})^t,
\end{equation*}
where the superscript $t$ stands for transposition.
We write $\angmom$ and $\angmom_\oplus$ for the orbital angular momenta of the asteroid and the Earth, respectively, and introduce their unit vectors
\begin{equation*}
    \hat{\angmom} = \frac{\angmom}{c}, \qquad
    \hat{\angmom}_\oplus = \frac{\angmom_\oplus}{c_\oplus},
\end{equation*}
where $c=|\angmom|$, $c_\oplus=|\angmom_\oplus|$, being $|\bm{x}|$ the Euclidean norm of a vector $\bm{x}$.

We also write $\br_i$ and $\bq_i$ for the heliocentric positions of the asteroid and the Earth at the three epochs $t_i$ ($i=1,2,3$), use $\rhovec_i = \br_i - \bq_i$ for the geocentric position of the asteroid, and introduce the unit vectors
\begin{equation*}
    \qhat_i = \frac{\bq_i}{q_i},\qquad \rhohat_i = \frac{\rhovec_i}{\rho_i},\qquad i=1,2,3,
\end{equation*}
where $q_i=|\bq_i|$, $\rho_i=|\rhovec_i|$. 
Finally, we denote the parameters\footnote{in \cite{mossotti} these are called {\em semiparametri}.} of the orbits of the asteroid and the Earth by $p$, $p_\oplus$.
They are defined by
\begin{equation*}
    p\kappa^2 = c^2,\qquad p_\oplus\kappa^2 = c_\oplus^2.
\end{equation*}

\subsection{Geometric relations}

With the purpose of writing Mossotti's equations in a compact form, let us introduce the matrices
\begin{equation}
    P = (\bm{\rho}_1\, |\, \bm{\rho}_2  \,|\, \bm{\rho}_3),\qquad
    Q = (\bm{q}_1\, |\, \bm{q}_2  \,|\, \bm{q}_3),\qquad
    R = P + Q = (\bm{r}_1\, |\, \bm{r}_2  \,|\, \bm{r}_3),
\label{eq:PQR}
\end{equation}
where $(\bm{x}_1 \,|\, \bm{x}_2 \,|\, \bm{x}_3)$ is the matrix whose columns are the vectors $\bm{x}_j$. 
The corresponding adjugate matrices are
\begin{equation*}
    \begin{split}
    \adj(P) &=
    \bigl(\bm{\rho}_2\times\bm{\rho}_3\,|\,
    \bm{\rho}_3\times\bm{\rho}_1\,|\,
    \bm{\rho}_1\times\bm{\rho}_2\bigr)^t,\cr
    \adj(Q) &=
    \bigl(\bm{q}_2\times\bm{q}_3\,|\,
    \bm{q}_3\times\bm{q}_1\,|\,
    \bm{q}_1\times\bm{q}_2\bigr)^t,\cr
    \adj(R) &=
    \bigl(\bm{r}_2\times\bm{r}_3\,|\,
    \bm{r}_3\times\bm{r}_1\,|\,
    \bm{r}_1\times\bm{r}_2\bigr)^t.\cr
    \end{split}
\end{equation*}
We recall the following property, which holds for any square matrix $M$:
\begin{equation}
    M\,\adj(M) = \adj(M)\,M = \det(M)\,I,
\label{adjdet}
\end{equation}
where $I$ is the identity matrix.

\begin{figure}[ht!] 
    \begin{center}
    \includegraphics[trim= 0mm 0mm 0mm 0mm, clip, width=0.72\textwidth]{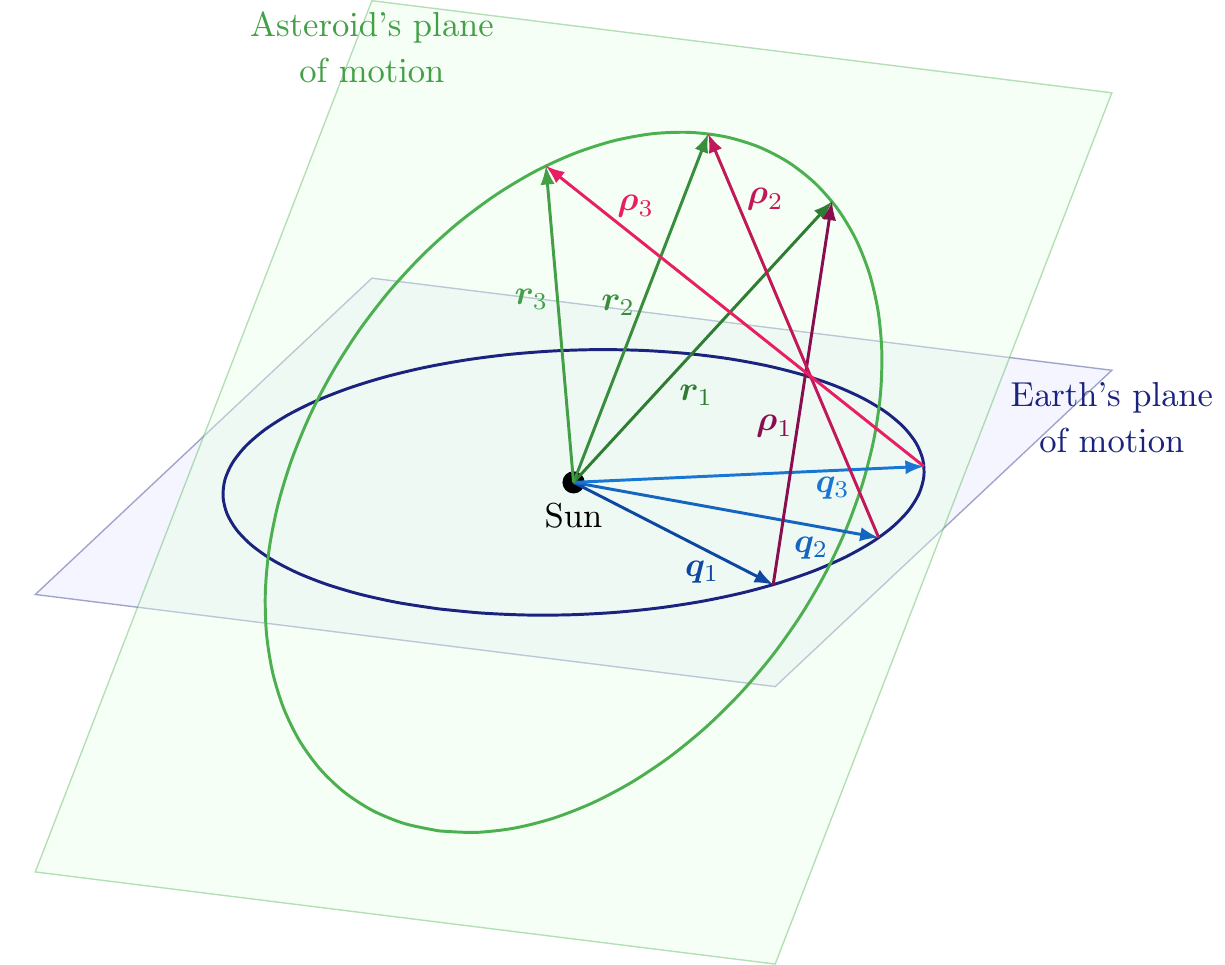}
   \end{center}
    \vspace{-1mm}
    \caption{Geometry of the three observations.}
    \label{fig:esquemaGeo}
\end{figure}

The rank of the matrices $Q$ and $R$ is 2, since each triplet $\{\bm{q}_1, \bm{q}_2,\bm{q}_3\}$ and $\{\bm{r}_1, \bm{r}_2, \bm{r}_3\}$ is made by coplanar vectors and we assume that
\begin{equation*}
    \bm{q}_i\times\bm{q}_j\neq \bm{0}, \qquad
    \bm{r}_i\times\bm{r}_j\neq \bm{0}, \qquad 1\leq i<j\leq 3,
\end{equation*}
see Figure~\ref{fig:esquemaGeo}. This implies
\begin{equation}
    Q\,\adj(Q)= 0, \qquad R\,\adj(R) = 0.
    \label{eq:qrLD}
\end{equation}
Since the angular momenta $\angmom$, $\angmom_\oplus$ are respectively orthogonal to the orbital planes of the asteroid and the Earth, we have
\begin{equation*}\label{eq:relCla}
    \angmom\cdot\br_i = \angmom\cdot(\bq_i+\rhovec_i)= 0,\quad
    \angmom_\oplus\cdot\bq_i = 0,
\end{equation*}
which lead us to
\begin{equation}\label{eq:relMomAng}
    \angmom\cdot\rhovec_i = (\angmom_\oplus-\angmom)\cdot\bq_i,
\end{equation}
for $i=1,2,3$. Then, we introduce the vectors
\begin{equation*}
    \bm{\tau}=(\tau_1,\,\tau_2,\,\tau_3)^t,\qquad \bm{T}=(T_1,\,T_2,\,T_3)^t,
\end{equation*}
with
\begin{subequations}
    \begin{equation}
        \tau_1\sqrt{p} = \erre_2\times\erre_3\cdot\hat{\angmom}, \hskip 1cm
        T_1\sqrt{p_\oplus} = \bq_2\times\bq_3\cdot\hat{\angmom}_\oplus,
    \end{equation}
    \begin{equation}
        \tau_2\sqrt{p} = \erre_3\times\erre_1\cdot\hat{\angmom},  \hskip 1cm
        T_2\sqrt{p_\oplus} = \bq_3\times\bq_1\cdot\hat{\angmom}_\oplus,
    \end{equation}
    \begin{equation}
        \tau_3\sqrt{p} = \erre_1\times\erre_2\cdot\hat{\angmom},  \hskip 1cm
        T_3\sqrt{p_\oplus} = \bq_1\times\bq_2\cdot\hat{\angmom}_\oplus.
    \end{equation}
    \label{eq:deftauTlong}
\end{subequations}
Noting that
\begin{equation} 
    \bm{\tau} = \frac{1}{\sqrt{p}} \adj(R)\hat{\bm{c}},\qquad
    \bm{T} = \frac{1}{\sqrt{p_\oplus}} \adj(Q)\hat{\bm{c}}_\oplus,
    \label{eq:deftauTmat}
\end{equation}
by \eqref{eq:qrLD} we have 
\begin{equation*}
    R\,\bm{\tau} = \bm{0},
    \qquad
    Q\,\bm{T} = \bm{0}.
\end{equation*}
Moreover, recalling that $R = Q + P$, we get
\begin{equation}\label{eq:91_new}
    P\bm{\tau} = Q(\bm{T}-\bm{\tau}).
\end{equation}
Since $(\bm{r}_i\times\bm{r}_j)\times\angmom = (\bm{q}_i\times\bm{q}_j)\times\angmom_\oplus = {\bm 0}$, we have
\begin{align}
    \left(\bm{r}_i\times\bm{r}_j\cdot\hat{\bm{c}}\right)\left(\hat{\bm{c}}\cdot\bm{\rho}_i\right)
    &= \bm{r}_i\times\bm{r}_j\cdot\bm{\rho}_i
    = \bm{q}_i\times\bm{q}_j \cdot\bm{\rho}_i - \bm{\rho}_i\times\bm{\rho}_j \cdot\bm{q}_i,
    \label{eq:georel1}\\
    (\bm{q}_i\times\bm{q}_j\cdot\hat{\angmom}_\oplus)(\hat{\angmom}_\oplus\cdot\bm{\rho}_i)
    &= \bm{q}_i\times\bm{q}_j\cdot\bm{\rho}_i.\label{eq:georel2}
\end{align}
\begin{remark}
    If the orbits of the Earth and the asteroid are almost coplanar, then the equations \eqref{eq:georel1}, \eqref{eq:georel2} are close to be degenerate. 
    This singularity is a common feature of orbit determination methods, where the value of the geodesic curvature of the observed arc plays an important role, see \cite[Chap. 9]{mg2010}.
\end{remark}
The last equality in \eqref{eq:georel1} is obtained by observing that
\begin{equation*}
    \bm{r}_i\times\bm{r}_j \cdot\bm{\rho}_i = \bm{q}_i\times\bm{r}_j
    \cdot\bm{\rho}_i = \bm{q}_i\times\bm{q}_j \cdot\bm{\rho}_i +
    \bm{q}_i\times\bm{\rho}_j \cdot\bm{\rho}_i.
\end{equation*}
By \eqref{eq:georel1}, \eqref{eq:georel2} and the definitions of the $\tau_k$, $T_k$ given in \eqref{eq:deftauTlong} we obtain
\begin{align}
    \varepsilon_{ijk}\frac{\tau_k}{\kappa}\bm{c}\cdot\bm{\rho}_i &= \bm{q}_i\times\bm{q}_j \cdot\bm{\rho}_i -
    \bm{\rho}_i\times\bm{\rho}_j \cdot\bm{q}_i,\label{eq:reltau}\\[1ex]
    \varepsilon_{ijk}\frac{T_k}{\kappa}\bm{c}_\oplus\cdot\bm{\rho}_i &= \bm{q}_i\times\bm{q}_j \cdot\bm{\rho}_i,
    \label{eq:relbigT}
\end{align}
where $\varepsilon_{ijk}$ denotes the Levi-Civita symbol, and the indexes $i,j,k$ vary so that all the 6 permutations of the set $\{1,2,3\}$ can be considered.

Subtracting \eqref{eq:reltau} from \eqref{eq:relbigT} we obtain
\begin{equation}
    \varepsilon_{ijk}\frac{1}{\kappa}(T_k\angmom_\oplus-\tau_k\angmom)\cdot\bm{\rho}_i = 
    \bm{\rho}_i\times\bm{\rho}_j \cdot\bm{q}_i.
    \label{eq:relTtau}
\end{equation}
Writing
\begin{equation*}
    T_k\bm{c}_\oplus- \tau_k\bm{c}
    = T_k\left(\bm{c}_\oplus-\bm{c}\right) +\left(T_k-\tau_k\right)\bm{c},
\end{equation*}
and using \eqref{eq:relMomAng}, we have
\begin{equation*}
    \begin{aligned}
    \left[T_k\left(\bm{c}_\oplus-\bm{c}\right)
      +\left(T_k-\tau_k\right)\bm{c}\right]\cdot\bm{\rho}_i &= 
    T_k\left(\bm{c}_\oplus-\bm{c}\right)\cdot\bm{\rho}_i
    +\left(T_k-\tau_k\right)\left(\bm{c}_\oplus-\bm{c}\right)\cdot\bm{q}_i \\
    &=  \left(\bm{c}_\oplus-\bm{c}\right)\cdot\left[\hat{\bm{\rho}}_i
      +\left(1-\frac{\tau_k}{T_k}\right)\frac{q_i}{\rho_i}\hat{\bm{q}}_i\right]T_k\rho_i.
    \end{aligned}
\end{equation*}

\noindent Therefore, introducing the coefficients
\begin{equation}
    \alpha_{ik} = \left(1-\frac{\tau_k}{T_k}\right)\frac{q_i}{\rho_i},
    \label{eq:alpha_def}
\end{equation}
and simplifying $\rho_i$ we can write relations \eqref{eq:relTtau} as
\begin{equation}
    \varepsilon_{ijk}\left(\bm{c}_\oplus-\bm{c}\right)\cdot
    \left(\hat{\bm{\rho}}_i+\alpha_{ik}\hat{\bm{q}}_i\right)T_k =
    \kappa\rho_j\hat{\bm{\rho}}_i\times\hat{\bm{\rho}}_j\cdot\bm{q}_i.
    \label{eq:relTtau_new}
\end{equation}

\subsection{Combining geometry of observations with two-body dynamics}

Since the time intervals $\theta_{12}$, $\theta_{23}$ are small compared to the orbital period,\footnote{ the observations of asteroids at Mossotti's epoch were no more than one per night, and the time interval between two of them covered a few days.}
we can consider Taylor's expansions of the position of the asteroid and the center of the Earth in their orbital planes as power series of $\theta_{23}$, $\theta_{31}$, $\theta_{12}$ centering the expansion at the intermediate epoch. 
Then, we neglect the terms depending on the powers of $\theta_{ij}$ greater than 3. 
In this way we obtain
\begin{equation}
    \bm{\tau} \simeq \bm{\theta} - \frac{1}{6r_2^3}\bm{\theta}^3, \quad\quad
    \bm{T} \simeq \bm{\theta} - \frac{1}{6q_2^3}\bm{\theta}^3,
    \label{eq:tau_T_new}
\end{equation}
where $r_2=|\br_2|$ and
\begin{equation*}
    \bm{\theta}^3 = \bm{\theta} \odot \bm{\theta} \odot \bm{\theta} = \left(
    \theta_{23}^3,\,\theta_{31}^3,\,\theta_{12}^3\right)^t,
\end{equation*}
with $\odot$ denoting the Hadamard product.\footnote{ if $\bm{a}=(a_1,a_2,a_3)^t$ and $\bm{b}=(b_1,b_2,b_3)^t$, then $\bm{a}\odot\bm{b} = (a_1b_1,a_2b_2,a_3b_3)^t$.}

Let us define
\begin{equation*}
    \bm{u}=(u_1,\,u_2,\,u_3)^t=\adj(\hat{P})Q{\bm{\theta}^3},
\end{equation*}
with $\hat{P} = (\hat{\bm{\rho}}_1\, |\, \hat{\bm{\rho}}_2 \,|\,
\hat{\bm{\rho}}_3)$.
Multiplying relation \eqref{eq:91_new} on the left by $\adj(\hat{P})$, and taking into account the approximations \eqref{eq:tau_T_new}, we get
\begin{equation}
    \begin{aligned}
        \adj(\hat{P})P\bm{\theta} & \simeq \adj(\hat{P})P\bm{\tau} = \adj(\hat{P})Q(\bm{T}-\bm{\tau})\\
        & \simeq \adj(\hat{P})Q\bm{\theta}^3\left(\frac{1}{6r_2^3} - \frac{1}{6q_2^3}\right)
        = \frac{1}{6}\left(\frac{1}{r_2^3} - \frac{1}{q_2^3}\right)\bm{u}.
  \end{aligned}
  \label{utheta}
\end{equation}
Recalling \eqref{adjdet}, and noting that
\begin{equation*}
\adj(\hat{P})P = \det(\hat{P})\textrm{diag}\{\rho_1,\rho_2,\rho_3\},
\end{equation*}
relation
\eqref{utheta} yields
\begin{equation}
    \det(\hat{P})\bm{\delta}\odot\bm{\theta} \simeq \frac{1}{6}\left(\frac{1}{r_2^3} - \frac{1}{q_2^3}\right) \bm{u},
    \label{eq:urel}
\end{equation}
with
\begin{equation*}
    \bm{\delta} = (\rho_1,\rho_2,\rho_3)^t.
\end{equation*}

\subsection{A linear equation involving $\angmom_\oplus-\angmom$}

Choosing $(i,j,k)=(1,2,3), (3,2,1)$ in \eqref{eq:relTtau_new} and eliminating $\kappa\rho_2$ from the two resulting equations, we obtain
\begin{equation}\label{eq:s2Geo}
    \begin{split}
    & \left(\hat{\bm{\rho}}_1\times\hat{\bm{\rho}}_2 \cdot\bm{q}_1\right)
    \left[\left(\bm{c}_\oplus-\bm{c}\right)\cdot
      \left(\hat{\bm{\rho}}_3 + \alpha_{31}\hat{\bm{q}}_3\right)\right]T_1
    =\\
    & \left(\hat{\bm{\rho}}_2\times\hat{\bm{\rho}}_3 \cdot\bm{q}_3\right)
    \left[\left(\bm{c}_\oplus-\bm{c}\right)\cdot
      \left(\hat{\bm{\rho}}_1 + \alpha_{13}\hat{\bm{q}}_1\right)\right]T_3.
    \end{split}
\end{equation}
In  this equation the only unknowns different from $\left(\bm{c}_\oplus-\bm{c}\right)$ are the coefficients $\alpha_{13}$ and $\alpha_{31}$.
Using the approximations of $\bm{\tau}$ and $\bm{T}$ given by
\eqref{eq:tau_T_new} in \eqref{eq:alpha_def} we have
\begin{equation}\label{eq:alph13}
    \alpha_{13} = \frac{q_1}{\rho_1}\frac{T_3-\tau_3}{T_3}
    \simeq
    \frac{q_1}{\rho_1T_3}
    \frac{\theta_{12}^3}{6}\left(\frac{1}{r_2^3}-\frac{1}{q_2^3}\right)
    \simeq
    \frac{q_1}{\rho_1}
    \frac{\theta_{12}^2}{6}\left(\frac{1}{r_2^3}-\frac{1}{q_2^3}\right),
\end{equation}
where we used $\frac{1}{T_3} = \frac{1}{\theta_{12}}(1+O(\theta_{12}))$.
In a similar way we obtain
\begin{equation}\label{eq:alph31}
    \alpha_{31} 
    \simeq
    \frac{q_3}{\rho_3}
    \frac{\theta_{23}^2}{6}\left(\frac{1}{r_2^3}-\frac{1}{q_2^3}\right).
\end{equation}
Inserting the approximations of $\rho_1, \rho_3$ given by \eqref{eq:urel} into \eqref{eq:alph13}, \eqref{eq:alph31} we can express $\alpha_{13}$, $\alpha_{31}$ with known quantities:
\begin{equation}\label{eq:alphas}
    \alpha_{13} \simeq \frac{\det(\hat{P})q_1{\theta_{12}^{2}\theta_{23}}}{u_1},
    \qquad
    \alpha_{31} \simeq \frac{\det(\hat{P})q_3{\theta_{23}^2\theta_{12}}}{u_3}.
\end{equation}
Defining the coefficients
\begin{equation*}
    a_{1} = \frac{q_2\left(\hat{\bm{\rho}}_2\times\hat{\bm{\rho}}_3 \cdot\bm{q}_3\right)}{T_1\sqrt{p_\oplus}}
    = \frac{\hat{\bm{\rho}}_2\times\hat{\bm{\rho}}_3 \cdot\hat{\bm{q}}_3}{\hat{\bm{q}}_2\times\hat{\bm{q}}_3
    \cdot\hat{\bm{c}}_\oplus},
    \qquad
    a_{3} = \frac{q_2\left(\hat{\bm{\rho}}_1\times\hat{\bm{\rho}}_2 \cdot\bm{q}_1\right)}{T_3\sqrt{p_\oplus}}
    = \frac{\hat{\bm{\rho}}_1\times\hat{\bm{\rho}}_2 \cdot\hat{\bm{q}}_1}{\hat{\bm{q}}_1\times\hat{\bm{q}}_2
    \cdot\hat{\bm{c}}_\oplus},
\end{equation*}
and the vectors
\begin{equation*}
    \bm{\gamma} = a_{1}\left(\hat{\bm{\rho}}_1 + \alpha_{13}\hat{\bm{q}}_1\right),\qquad
    \bm{\varphi} = a_{3}\left(\hat{\bm{\rho}}_3 + \alpha_{31}\hat{\bm{q}}_3\right),
\end{equation*}
equation \eqref{eq:s2Geo} becomes
\begin{equation}
    (\bm{\gamma}-\bm{\varphi})\cdot\left(\bm{c}_\oplus-\bm{c}\right) = 0.
  \label{eq:slingeo}
\end{equation}

\begin{remark}
    Choosing the index pair $\{(2,3,1),(1,3,2)\}$ or $\{(3,1,2),(2,1,3)\}$ in \eqref{eq:relTtau_new} we can obtain two additional equations analogous to \eqref{eq:slingeo}. 
    However, with the employed approximation of $\bm{T}$ and $\bm{\tau}$, the three equations are just the same (see \cite[Sect. 28]{mossotti}).
\end{remark}

\subsection{Mossotti's equations for $\angmom_\oplus-\angmom$}
\label{s:mosseqs}

With the aim of writing two independent linear equations, all the four observations are used. 
If we consider two different choices of the three observations, out of the available four, we obtain the system
\begin{equation}
    \left\{
    \begin{aligned}
        &(\bm{\gamma}_1-\bm{\varphi}_1)\cdot\left(\bm{c}_\oplus-\bm{c}\right) = 0\\
        &(\bm{\gamma}_2-\bm{\varphi}_2)\cdot\left(\bm{c}_\oplus-\bm{c}\right) = 0,
    \end{aligned}
    \right.
    \label{system_geo}
\end{equation}
where the subscripts $1$, $2$ of $\bm{\gamma}$ and $\bm{\varphi}$ refer to the two triplets of observations.
Set
\begin{equation*}
    \bm{w} = (\bm{\gamma}_1-\bm{\varphi}_1)\times(\bm{\gamma}_2-\bm{\varphi}_2)
\end{equation*}
and assume $\bm{w}\neq\bm{0}$.
Then the general solution of \eqref{system_geo} has the form
\begin{equation}
    \bm{c}_\oplus - \bm{c} = \lambda\bm{w},
    \label{Cmcdir}
\end{equation}
with $\lambda\in\mathbb{R}$, giving the direction of $\bm{c}_\oplus - \bm{c}$.
\begin{remark}
    If $\bm{\gamma}_1-\bm{\varphi}_1$ and $\bm{\gamma}_2-\bm{\varphi}_2$ are almost parallel, then system \eqref{system_geo} is almost degenerate: we can try to avoid this singularity by choosing other triplets of observations.
\end{remark}
In order to constrain the values of $\lambda$ we proceed as follow.
Choosing $(i,j,k) = (1,2,3)$ in equation \eqref{eq:relTtau_new}, we have
\begin{equation}
    \frac{a_3 c_\oplus\rho_2}{q_2} = (\bm{c}_\oplus-\bm{c})\cdot
    \left(\hat{\bm{\rho}}_1+\alpha_{13}\hat{\bm{q}}_1\right).
\label{geomeq}
\end{equation}
Inserting the expression \eqref{Cmcdir} of the general solution in \eqref{geomeq} and \eqref{eq:relMomAng} with $i=2$ we obtain, respectively, 
\begin{equation}\label{eq:rho2norm}
    \rho_2 = \lambda\frac{q_2}{a_1a_3c_\oplus}\bm{w}\cdot\bm{\gamma}_1,
\end{equation}
and
\begin{equation}\label{eq:geomRel}
    \lambda\bm{w}\cdot\bm{q}_2 = (\bm{c}_\oplus - \lambda\bm{w})\cdot\bm{\rho}_2
    = (\bm{c}_\oplus - \lambda\bm{w})\cdot\hat{\bm{\rho}}_2\rho_2.
\end{equation}
Substituting the expression \eqref{eq:rho2norm} of $\rho_2$ in \eqref{eq:geomRel} yields a quadratic equation in $\lambda$, which is here the only unknown:
\begin{equation*}
    \lambda\bm{w}\cdot\bm{q}_2 = \lambda\frac{q_2}{a_1a_3 c_\oplus}\left(\bm{w}\cdot\bm{\gamma}_1\right)\big[(\bm{c}_\oplus
    - \lambda\bm{w})\cdot\hat{\bm{\rho}}_2\big].
\end{equation*}
This equation can be written as
\begin{equation}\label{eq:slgeo}
    (\bm{w}\cdot\bm{\gamma}_1)(\bm{w}\cdot\hat{\bm{\rho}}_2) \lambda^2 + 
    \big[a_1a_3c_\oplus(\bm{w}\cdot\hat{\bm{q}}_2)-(\bm{w}\cdot\bm{\gamma}_1)
    (\bm{c}_\oplus\cdot\hat{\bm{\rho}}_2)\big]\lambda = 0,
\end{equation}
whose  solutions are
\begin{subequations}
    \begin{equation}
        \lambda = 0,
        \label{lambdazero}
    \end{equation}
    \begin{equation}
        \lambda = \frac{(\bm{w}\cdot\bm{\gamma}_1)(\bm{c}_\oplus\cdot\hat{\bm{\rho}}_2) -
        a_1a_3c_\oplus(\bm{w}\cdot\hat{\bm{q}}_2)}{(\bm{w}\cdot\bm{\gamma}_1)(\bm{w}\cdot\hat{\bm{\rho}}_2)}.
        \label{lambdasol}
    \end{equation}
\end{subequations}
Substituting these expressions in \eqref{Cmcdir} gives two possible values of the angular momentum $\angmom$.
The solution \eqref{lambdazero} yields $\angmom_\oplus=\angmom$, and is usually discarded, so that \eqref{lambdasol} is regarded as the only solution.
In this way the equations of Mossotti's method can be considered linear.

\section{Topocentric method}
\label{s:mossotti_top}

\begin{figure}[t!] 
    \begin{center}
        \begin{tikzpicture}[domain=0:2, scale = 0.7]
            \draw (0,0) circle (0.5) (0,-0.5) node[below]{Sun};
            \draw (6,0) circle (1) (6,-1) node[below]{Earth} (2,5) node[above]{Asteroid};
            \draw[-latex, line width = 0.30mm](0,0) -- (6,0) node[midway,below]{$\bm{q}_\oplus$};
            \draw[-latex, line width = 0.30mm](0,0) -- (6.34202014332,0.93969262078) node[midway,above]{$\bm{q}$};
            \draw[-latex, line width = 0.30mm](6,0) -- (6.34202014332,0.93969262078) node[midway,right]{$\pobs$};
            \draw[-latex, line width = 0.30mm](0,0) -- (2,5) node[midway,left]{$\bm{r}$};
            \draw[-latex, line width = 0.30mm](6.34202014332,0.93969262078) -- (2,5) node[midway,above]{$\rhovec$};
            \draw[-latex, line width = 0.30mm](6,0) -- (2,5) node[midway,left]{$\rhovec_{\rm geo}$};
        \end{tikzpicture}
    \end{center}
    \vspace{-8mm}
    \caption{Geocentric and topocentric point of view.}
    \label{fig:esquema}
\end{figure}
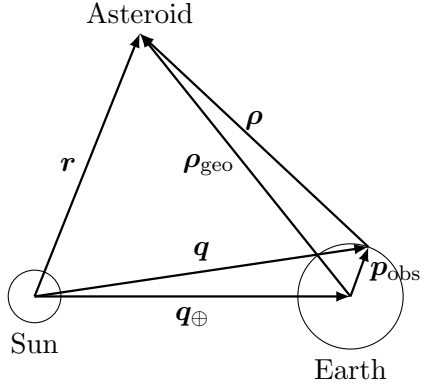

We first introduce some notation. Let us define $\bq_\oplus$, $\pobs$
as the heliocentric position of the Earth center, and the geocentric
position of the observer, respectively. The heliocentric positions of
the observer and asteroid are
\begin{equation*}
    \begin{aligned}
    & \bq=\bq_\oplus+\pobs,\cr
    & \erre = \bq + \rhovec = \bq_\oplus + \rhovec_{\rm geo},\cr
    \end{aligned}
\end{equation*}
where $\rhovec$, $\rhovec_{\rm geo}$ are the topocentric and geocentric positions of the asteroid, respectively (see Figure~\ref{fig:esquema}).

\subsection{Geometric relations}

As in the geocentric case, we select three observations of the asteroid out of the available four, and introduce the matrices $P$, $Q$, $R$ as in \eqref{eq:PQR}, but with a different interpretation for the vectors $\bm{\rho}$ and $\bm{q}$: here $\bm{\rho}$ represents the topocentric position of the asteroid, and $\bm{q}$ gives the heliocentric position of the observer.
Moreover, we introduce the matrices
\begin{equation*}
    Q_\oplus = {\bigl(\bm{q}_{\oplus,1}\, |\,   
                      \bm{q}_{\oplus,2}\, |\, 
                      \bm{q}_{\oplus,3}\bigr)}, \qquad
    P_\text{obs} = {\bigl(\bm{p}_{\text{obs},1}\, |\,   
                          \bm{p}_{\text{obs},2}\, |\, 
                          \bm{p}_{\text{obs},3}\bigr)}.
\end{equation*}
We recall the geometrical relations
\begin{equation*}\label{eq:relClaTop}
    \angmom\cdot\br_i = \angmom\cdot(\bq_i+\rhovec_i)= 0,\quad \angmom_\oplus\cdot\bq_{\oplus,i} =
    \angmom_\oplus\cdot(\bq_i - \bm{p}_{\text{obs},i}) = 0,
\end{equation*}
that lead us to
\begin{equation}\label{eq:relMomAngTop}
    \angmom\cdot\rhovec_i = (\angmom_\oplus-\angmom)\cdot\bq_i - \angmom_\oplus\cdot\bm{p}_{\text{obs},i},
\end{equation}
for $i=1,2,3$.
We also introduce the quantities 
\begin{equation}
    \bm{\tau} = \frac{1}{\sqrt{p}} \adj(R)\hat{\bm{c}},
    \qquad
    \bm{T} = \frac{1}{\sqrt{p_\oplus}} \adj(Q_\oplus)\hat{\bm{c}}_\oplus,
    \label{Ttau_top}
\end{equation}
which are the same as in \eqref{eq:deftauTmat}, and define the matrix
\begin{equation}
    \mathcal{C} = \left[\adj(Q)-\adj(Q_\oplus)\right]\hat{P}.
    \label{eq:matrixC_top}
\end{equation}
Like in the geocentric case, we have the relations
\begin{align}
    \varepsilon_{ijk}\frac{\tau_k}{\kappa}\bm{c}\cdot\bm{\rho}_i &= \bm{q}_i\times\bm{q}_j \cdot\bm{\rho}_i -
    \bm{\rho}_i\times\bm{\rho}_j \cdot\bm{q}_i,\label{eq:reltau_top}\\[1ex]
    \varepsilon_{ijk}\frac{T_k}{\kappa}\bm{c}_\oplus\cdot\bm{\rho}_i + {\cal C}_{ki}\rho_i &= \bm{q}_i\times\bm{q}_j \cdot\bm{\rho}_i,
    \label{eq:relbigT_top}
\end{align}
where $(i,j,k)$ is varied  so  that  all  the  6 permutations of the set $\{1,2,3\}$ are considered, and $\mathcal{C}_{ki}$ denotes the element of the $k$-th row and $i$-th column of the matrix $\mathcal{C}$.

Subtracting \eqref{eq:reltau_top} from \eqref{eq:relbigT_top} we obtain
\begin{equation}
    \varepsilon_{ijk}\frac{1}{\kappa}(T_k\angmom_\oplus-\tau_k\angmom)\cdot\bm{\rho}_i + {\cal C}_{ki}\rho_i= 
    \bm{\rho}_i\times\bm{\rho}_j \cdot\bm{q}_i,
\label{eq:relTtau_top}
\end{equation}
and following the same procedure as in the geocentric case we can write \eqref{eq:relTtau_top} as 
\begin{equation}
    \varepsilon_{ijk}\left[\left(\bm{c}_\oplus-\bm{c}\right)\cdot
    \left(\hat{\bm{\rho}}_i+\alpha_{ik}\hat{\bm{q}}_i\right) + \frac{\kappa\mathcal{C}_{ki}}{T_k}-\frac{\alpha_{ik}}{q_i}\bm{c}_\oplus\cdot\bm{p}_{\text{obs},i}\right]T_k =
    \kappa\rho_j\hat{\bm{\rho}}_i\times\hat{\bm{\rho}}_j\cdot\bm{q}_i,
\label{eq:relTtau_top_new}
\end{equation}
where the coefficients $\alpha_{ik}$ are defined as in \eqref{eq:alpha_def}, with the new interpretation for $\rho_i$ and $q_i$.

Using the same approximations as in \eqref{eq:tau_T_new}, we also note that
\begin{equation}
    \begin{aligned}
        \adj(\hat{P})P\bm{\theta} & \simeq \adj(\hat{P})P\bm{\tau} = \adj(\hat{P})Q_\oplus(\bm{T}-\bm{\tau}) - \adj(\hat{P})P_\text{obs}\bm{\tau}\\
        &\simeq \frac{1}{6}\bigg(\frac{1}{r_2^3} - \frac{1}{q_{\oplus,2}^3}\bigg)\bm{u}
        - \adj(\hat{P})P_\text{obs}\bm{\theta},
    \end{aligned}
    \label{eq:utheta_top}
\end{equation}
where $q_{\oplus,2} = |\bm{q}_{\oplus,2}|$ and
\begin{equation*}
\bm{u} = (u_1,u_2,u_3)^t = \adj(\hat{P})Q_\oplus \bm{\theta}^3.
\end{equation*}
The presence of the term $\adj(\hat{P})P_\text{obs}\bm{\theta}$ in \eqref{eq:utheta_top} prevents us from making the same simplification that allowed to express the $\alpha_{ik}$ as functions of known quantities.  
Noting that
\begin{equation*}
    \adj(\hat{P})P_\text{obs} = \mathcal{O}(p_\text{obs}),
\end{equation*}
where $p_{\text{obs}} = |\bm{p}_{\text{obs}}|$, we neglect this term and obtain
\begin{equation*}
    \det(\hat{P})\bm{\delta}\odot\bm{\theta} \simeq \frac{1}{6}\left(\frac{1}{r_2^3}
    - \frac{1}{q_{\oplus,2}^3}\right) \bm{u},
\end{equation*}
with
\begin{equation*}
\bm{\delta} = (\rho_1,\rho_2,\rho_3)^t.
\end{equation*}
As a consequence, the expressions for $\alpha_{13}$ and $\alpha_{31}$ given in \eqref{eq:alphas} can still be used, with the new interpretation for the vectors $\bm{\rho}$ and $\bm{q}$.

\subsection{A linear equation involving $\angmom_\oplus-\angmom$}

Choosing $(i,j,k)=(1,2,3), (3,2,1)$ in \eqref{eq:relTtau_top_new} and eliminating $\kappa\rho_2$ from the resulting equations, we obtain
\begin{equation}\label{eq:s2Top}
    \begin{aligned}
        & a_3 \left[\left(\bm{c}_\oplus-\bm{c}\right)\cdot
        \left(\hat{\bm{\rho}}_3
        +\alpha_{31}\hat{\bm{q}}_3\right) + \frac{\kappa\mathcal{C}_{13} 
        }{T_1}
        -\frac{\alpha_{31}}{q_3}\bm{c}_\oplus\cdot\bm{p}_{\text{obs},3}
        \right]
        =\\
        & a_1\left[(\bm{c}_\oplus-\bm{c})\cdot
        \left(\hat{\bm{\rho}}_1
        +\alpha_{13}\hat{\bm{q}}_1\right) + \frac{\kappa\mathcal{C}_{31}}{T_3}
        -\frac{\alpha_{13}}{q_1}\bm{c}_\oplus\cdot\bm{p}_{\text{obs},1}
        \right],
    \end{aligned}
\end{equation}
with
\begin{equation}
    a_{1} = \frac{\left(
    \hat{\bm{\rho}}_2\times\hat{\bm{\rho}}_3 \cdot\bm{q}_3
    \right){q}_2}{T_1\sqrt{p_\oplus}}
    = 
    \frac{\left(
    \hat{\bm{\rho}}_2\times\hat{\bm{\rho}}_3 \cdot\bm{q}_3
    \right){q}_2}{\bm{q}_{\oplus,2}\times\bm{q}_{\oplus,3} \cdot\hat{\bm{c}}_\oplus}
    , \quad
    a_{3} = \frac{\left(
    \hat{\bm{\rho}}_1\times\hat{\bm{\rho}}_2 \cdot\bm{q}_1
    \right){q}_2}{T_3\sqrt{p_\oplus}}
    = \frac{\left(
    \hat{\bm{\rho}}_1\times\hat{\bm{\rho}}_2 \cdot\bm{q}_1
    \right){q}_2}{\bm{q}_{\oplus,1}\times\bm{q}_{\oplus,2} \cdot\hat{\bm{c}}_\oplus}.
\label{eq:ajtop}
\end{equation}
Like in the geocentric case, defining the vectors
\begin{equation*}
    \bm{\gamma} = a_{1}\left(\hat{\bm{\rho}}_1 + \alpha_{13}\hat{\bm{q}}_1\right),\qquad
    \bm{\varphi} = a_{3}\left(\hat{\bm{\rho}}_3 + \alpha_{31}\hat{\bm{q}}_3\right),
\end{equation*}
we can write equation \eqref{eq:s2Top} as
\begin{equation}\label{eq:slintop}
    (\bm{\gamma}-\bm{\varphi})\cdot\left(\bm{c}_\oplus-\bm{c}\right) = D,
\end{equation}
with
\begin{equation*}
\begin{aligned}
    D &= \kappa\left(\frac{a_3\mathcal{C}_{13}}{T_1} - \frac{a_1\mathcal{C}_{31}}{T_3}\right)
    +\left(
    \frac{a_1\alpha_{13}}{q_1}\bm{p}_{\text{obs},1} 
    -
     \frac{a_3\alpha_{31}}{q_3}\bm{p}_{\text{obs},3} 
    \right) \cdot \bm{c}_\oplus.
\end{aligned}
\end{equation*}

\begin{remark}
    We can add to \eqref{eq:slintop} two equations choosing the index pairs $\{(2,3,1),(1,3,2)\}$ and $\{(3,1,2),(2,1,3)\}$ in     \eqref{eq:relTtau_top_new}.  
    In this way, we can write a linear system of three equations for $\angmom_\oplus-\angmom$ using only three observations. 
    However, we expect that the matrix of this system is ill-conditioned because $Q = Q_\oplus + \mathcal{O}(p_\obs)$, so that the vectors $\bq_1, \bq_2, \bq_3$ are almost coplanar.
\end{remark}

\subsection{Equations of the topocentric method}
\label{s:top}

Following Section~\ref{s:mosseqs}, if we consider two different choices of the three observations, we obtain the system
\begin{equation}
    \left\{
    \begin{aligned}
        &(\bm{\gamma}_1-\bm{\varphi}_1)\cdot\left(\bm{c}_\oplus-\bm{c}\right) = D_1\\
        &(\bm{\gamma}_2-\bm{\varphi}_2)\cdot\left(\bm{c}_\oplus-\bm{c}\right) = D_2.
    \end{aligned}
    \right.
\label{eq:linsisTop}
\end{equation}
Set
\begin{equation*}
    \bm{w} = (\bm{\gamma}_1-\bm{\varphi}_1)\times(\bm{\gamma}_2-\bm{\varphi}_2),
\end{equation*}
and assume $\bm{w}\neq\bm{0}$.
Then the general solution of \eqref{eq:linsisTop} takes the form
\begin{equation}
    \bm{c}_\oplus - \bm{c} = \lambda\bm{w} + \bm{g},
\label{gensol_top}
\end{equation}
where $\lambda\in\mathbb{R}$, and $\bm{g}$ is a particular solution of \eqref{eq:linsisTop}, e.g. the one fulfilling $\bm{g}\cdot\bm{w}=0$. 

In order to constrain the values of $\lambda$ we proceed as follow.
Note that we can write \eqref{eq:relTtau_top_new} with $(i,j,k) = (1,2,3)$ as
\begin{equation}
    \frac{a_3 c_\oplus\rho_2}{q_2} = \left(\bm{c}_\oplus-\bm{c}\right)\cdot
        \left(\hat{\bm{\rho}}_1
        +\alpha_{13}\hat{\bm{q}}_1\right) + \frac{\kappa\mathcal{C}_{31}}{T_3} -\frac{\alpha_{13}}{q_1}\bm{c}_\oplus\cdot\bm{p}_{\text{obs},1}.
\label{thirdeq_top}
\end{equation}
Inserting the general solution \eqref{gensol_top} into \eqref{thirdeq_top} and \eqref{eq:relMomAngTop} {with $i=2$} we obtain
\begin{equation}
    {\rho}_2 = \frac{1}{b}(\lambda\bm{w}+\bm{g})\cdot\bm{\gamma}_1
    + f,
    \label{eq:rho2normTop}
\end{equation}
where
\begin{equation}
    b = \frac{a_1a_3c_\oplus}{q_2}, \qquad
    f=\frac{q_2}{a_3c_\oplus}\left(\frac{\kappa\mathcal{C}_{31}}{T_3} -\frac{\alpha_{13}}{q_1}\bm{c}_\oplus\cdot\bm{p}_{\text{obs},1} \right),
\label{bieffe}
\end{equation}
and
\begin{equation}
    \left(\angmom_\oplus - \lambda\bm{w} - \bm{g}\right)\cdot\hat{\rhovec}_2\,\rho_2
    = (\lambda\bm{w} + \bm{g})\cdot\bq_2
    - \angmom_\oplus \cdot \bm{p}_{\text{obs},2}.
    \label{eq:geomRelTop}
\end{equation}
Substituting the expression \eqref{eq:rho2normTop} of $\rho_2$ in \eqref{eq:geomRelTop} we get
\begin{equation*}
    \left(\angmom_\oplus - \lambda\bm{w} - \bm{g}\right)\cdot\hat{\rhovec}_2\left[ 
    \frac{1}{b}(\lambda\bm{w}+\bm{g})\cdot\bm{\gamma}_1 + f
    \right]= (\lambda\bm{w} + \bm{g})\cdot\bq_2 - \angmom_\oplus\cdot\bm{p}_{\text{obs},2},
\end{equation*}
which can be written as
\begin{equation}\label{eq:sctop}
    \begin{split}
    & (\bm{w}\cdot{\bm{\gamma}}_1)(\bm{w}\cdot\hat{\bm{\rho}}_2)\lambda^2
    +\bigl[(\bm{w}\cdot{\bm{q}}_2)b
    - (\bm{w}\cdot\bm{\gamma}_1)(\bm{c}_\oplus-\bm{g})\cdot\hat{\bm{\rho}}_2
    + (\bm{w}\cdot{\hat{\bm\rho}}_2)(\bm{g}\cdot\bm{\gamma}_1
    + b f)\bigr]\lambda\\[1ex]
    & + b\left(\bm{g}\cdot{\bm{q}_2}-\bm{c}_\oplus\cdot\bm{p}_{\rm{obs},2}\right)
    - \left(\bm{g}\cdot\bm{\gamma}_1+ b f\right)(\bm{c}_\oplus-\bm{g})\cdot\hat{\bm{\rho}}_2 = 0.
    \end{split}
\end{equation}
Equation \eqref{eq:sctop} can be compared with equation \eqref{eq:slgeo}.  
It is worth noting that in the topocentric formulation we do not have the solution $\lambda= 0$ as in Mossotti's original method, so that this formulation leads to a quadratic equation.

\begin{remark}
    In the topocentric case we could add a third linear equation to system \eqref{eq:linsisTop} by choosing three different triplets of observations, out of the available four.  
    However, we expect that also in this case the system is ill-conditioned, because the vectors $\bq_1,\bq_2,\bq_3,\bq_4$ are almost coplanar.
\end{remark}

\section{Mossotti's method for space debris}
\label{s:mossotti_deb}

We can follow the same scheme introduced in Section~\ref{s:mossotti_top} for the computation of the orbits of space debris, assuming that the Earth is spherical and rotates with uniform angular velocity.  
In this case we use the rescaled time
\begin{equation*}
    \theta = \kappa_\oplus t,
\end{equation*}
with $\kappa_\oplus = \sqrt{Gm_\oplus}$.
\begin{figure}
    \begin{center}
    \includegraphics[width=12cm]{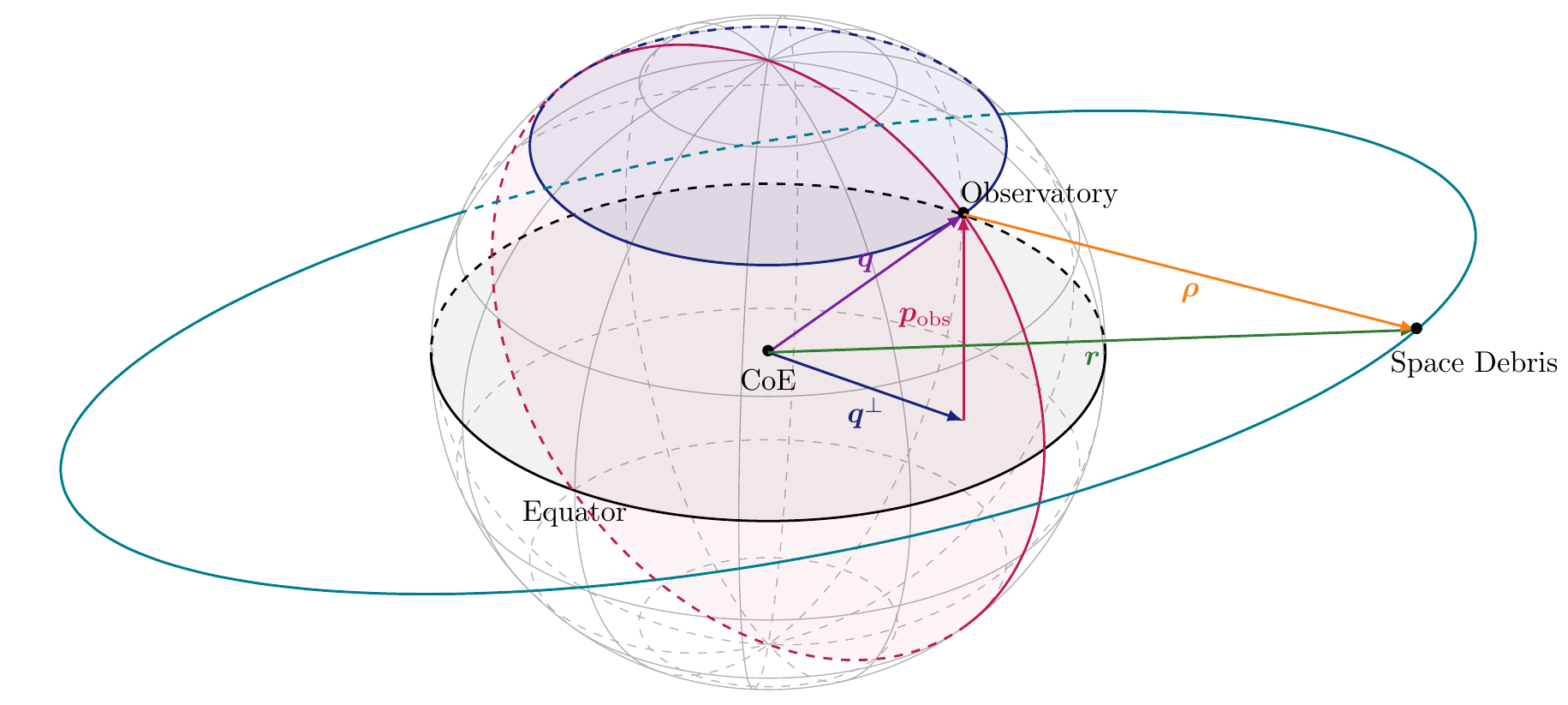}
    \end{center}
    \caption{Geometry of observations of space debris.}
    \label{fig:debris}
\end{figure}
Here the vector $\bq$ represents the geocentric positions of the observer, $\br$ and $\rhovec = \br - \bq$ give the geocentric and topocentric positions of the debris, and $\angmom=\erre\times\erredot$ is its orbital angular momentum.  
We consider the orthogonal decomposition
\begin{equation*}
    \bq = \bqperp + \pobs,
\end{equation*}
with
\begin{equation*}
    \bqperp = (\etre\times\bq)\times\etre, \qquad \pobs = (\bq\cdot\etre)\etre,
\end{equation*}
where $\etre$ is the unit vector of the Earth rotation axis, see Figure~\ref{fig:debris}.
Moreover, we introduce the vector
\begin{equation*}
    \angmom_\obs = \bqperp\times\dbqperp= \bqperp\times\dbq,
\end{equation*}
where the last equality holds because $\pobs$ is constant.

We can write a quadratic equation analogous to \eqref{eq:sctop} simply by substituting the vectors $\angmom_\oplus$, $\bq_\oplus$ with $\angmom_\obs$, $\bqperp$, and the parameter $p_\oplus$ with $|\bqperp|$.

\section{Numerical tests}
\label{s:numtests}

In this section we test the performance of Mossotti's original geocentric method (see Section \ref{s:mossotti_geo}) and its topocentric version introduced in Section \ref{s:mossotti_top}. 
In the following, we denote the former by M$_{geo}$ and the latter by M$_{top}$.

In Sections \ref{s:noerror}, \ref{s:error} we compare M$_{geo}$ with M$_{top}$ using simulated observations (right ascension and declination) computed for the site of the Pan-STARRS1 telescope, mount Haleakala, Hawaii, without taking into account observability conditions, i.e. the asteroids are not necessarily visible in the night sky. 
Moreover, we assume that the four observations given in input to M$_{geo}$ and M$_{top}$ are equally spaced in time. 
The time interval $\Delta t$ between two consecutive observations is varied in the two intervals:
\begin{equation}
    I_1=[15,200]\text{ minutes,}\qquad I_2=[0.25,100]\text{ days.}
    \label{I12}
\end{equation}
The comparison is based on the computation of the angular momentum vector ${\bf c}$ and the following related quantities: its magnitude $c = |{\bf c}|$ and direction $\hat{{\bf c}}={\bf c}/c$, the longitude of the (ascending) node $\Omega$, and the inclination $i$. 
We denote by $x_t$ the \emph{true} value of the quantity $x$, and by $x_{geo}$, $x_{top}$ the values of $x$ computed by M$_{geo}$, M$_{top}$, respectively. 
Note that while M$_{geo}$ always produces one solution for ${\bf c}$, the method M$_{top}$ can give two solutions. 
If this is the case, we select the one for which $|{\bf c}_t-{\bf c}_{top}|$ is smaller.

In Section \ref{s:synthetic-survey-tests} the method M$_{top}$ is compared to Gauss's method for initial orbit determination.
Synthetic data have been obtained that take into account the observability conditions and the expected real cadence of the observations from the Vera C. Rubin Observatory, which is currently under construction in Chile. 
With respect to the previous tests, we remark that for any set of four observations the time interval $\Delta t$ is not constant.

\subsection{Tests without astrometric errors}
\label{s:noerror}

We consider simulated observations of the asteroid Vesta without astrometric error. 
The time $\Delta t$ between two consecutive observations is varied in the two intervals $I_1$, $I_2$ specified in \eqref{I12}. 
Given $\Delta t$, we select 10$^5$ different initial epochs in a random way, and for each of them we generate four observations. 
For this purpose, we use the orbit of Vesta at the epoch 59200 MJD from the AstDyS-2 website\footnote{https://newton.spacedys.com/astdys/, last access February 13, 2021.} and propagate it to the desired epochs assuming Keplerian motion. 
The angular momentum vector defined by the Keplerian orbit is assumed to be the true solution (${\bf c}_t$). 
Finally, we apply M$_{geo}$, M$_{top}$ to each set of four observations.

\begin{figure}[ht!]
    \centering 
    \includegraphics[trim= 0mm 12mm 0mm 0mm, clip, width=0.48\textwidth]{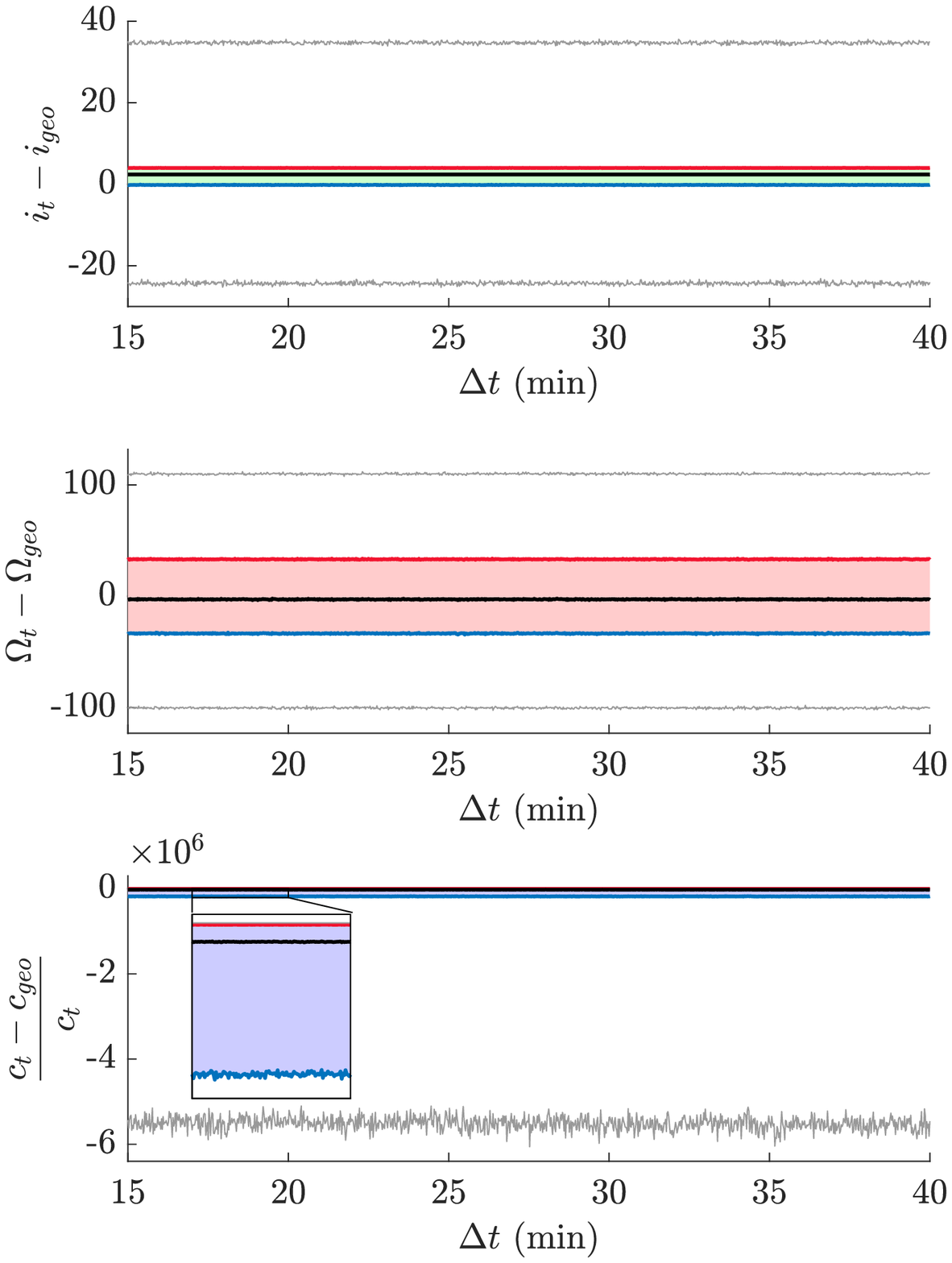}
    \includegraphics[trim= 0mm 12mm 0mm 0mm, clip, width=0.48\textwidth]{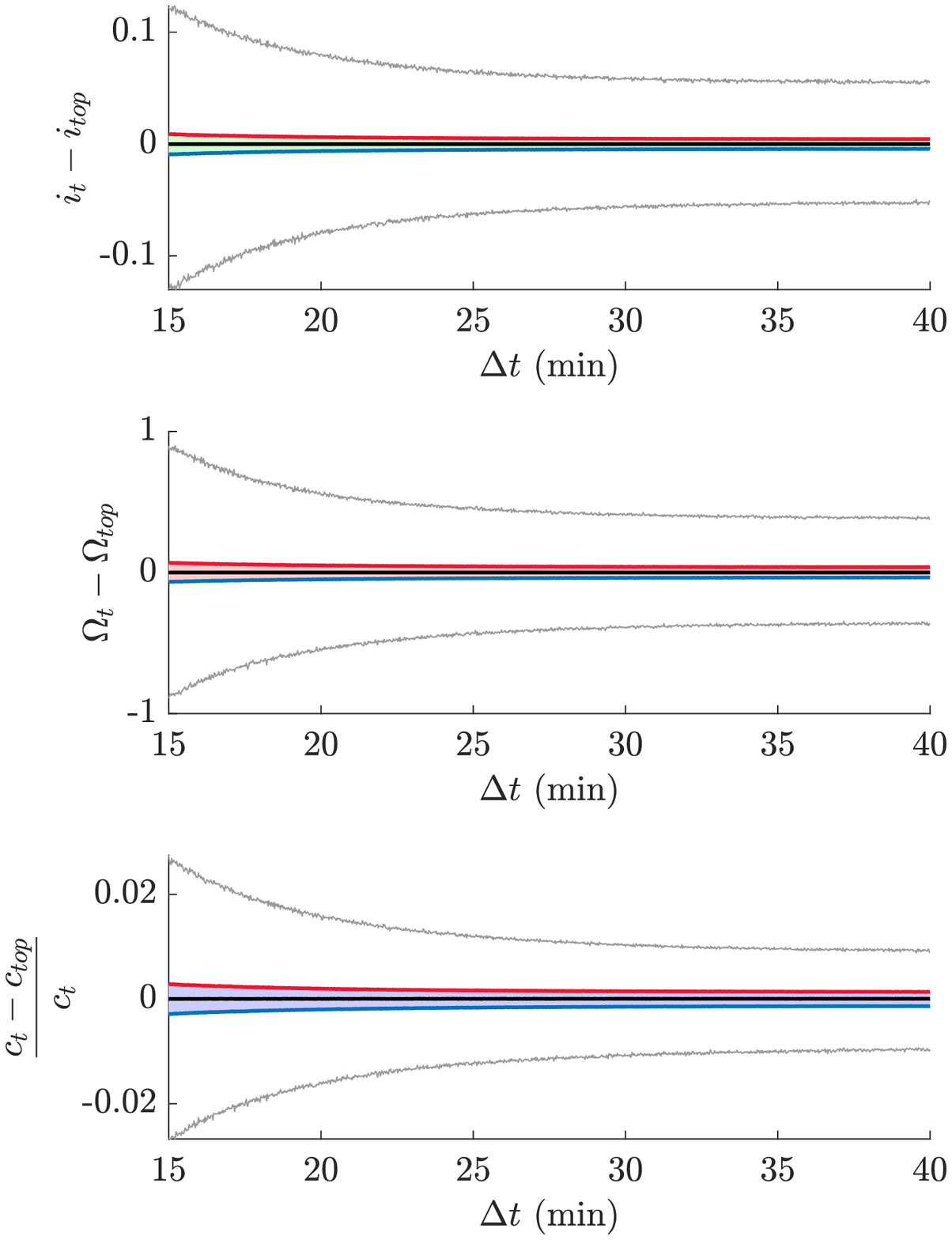}
    \caption{Statistics related to the errors in the inclination $i$, longitude of the node $\Omega$, and magnitude of the angular momentum vector $c$, obtained by M$_{geo}$ (\emph{left}) and M$_{top}$ (\emph{right}) as functions of $\Delta t$ in the case of observations of Vesta without astrometric error. 
    In particular, we show the median (black line), the 1st and 3rd quartiles (blue and red lines), and the 5th and 95th percentiles (gray lines). 
    The angles are in degrees { and the $\Delta t$ step is 1.5 seconds.}}
    \label{fig:VestaNoErr1}
\end{figure}

Let us first consider the case of short arcs of observation. 
For each $\Delta t$ selected in $I_1$ we compute the differences between the true values of the inclination ($i_t$), longitude of the node ($\Omega_t$), magnitude of the angular momentum vector ($c_t$) of Vesta, and the values obtained by either M$_{geo}$ or M$_{top}$. 
Some relevant statistical quantities related to these errors are shown in Figure \ref{fig:VestaNoErr1} as functions of $\Delta t$. We note that the topocentric version of Mossotti's method provides much better results than the original method. 
In particular, we observe that the performance of M$_{top}$ improves as $\Delta t$ increases, and it stabilizes when the time interval is about 30 minutes. 
It is remarkable that for $\Delta t$ larger than 25 minutes the error in the inclination is smaller than 0.01 degrees for the solutions that fall within the 1st and 3rd quartile and it is smaller than 0.1 degrees for the solutions that fall within the 5th and 95th percentile.

\begin{figure}[ht!]
    \centering
    \includegraphics[trim= 0mm 0mm 0mm 0mm, clip, width=0.97\textwidth]{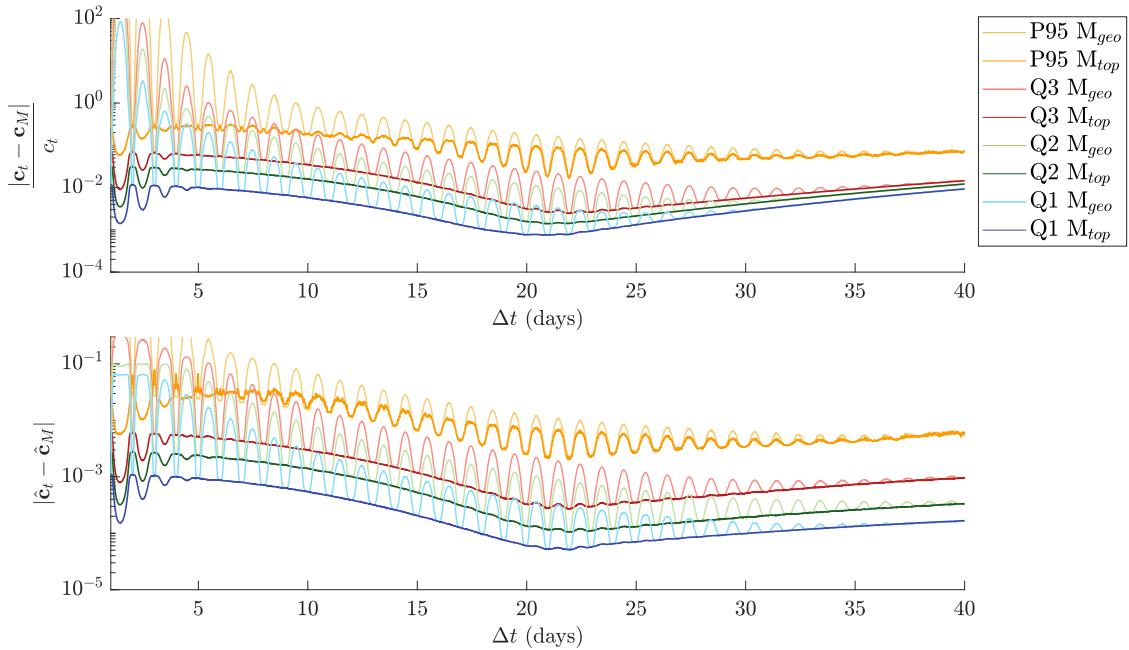}
    \caption{Statistics related to the errors in the angular momentum vector ${\bf c}$ and its direction $\hat{\bf c}$, obtained by M$_{geo}$ and M$_{top}$ as functions of $\Delta t$ in the case of observations of Vesta without astrometric error. 
    In particular, we show the median (Q2), the 1st and 3rd quartiles (Q1, Q3), and the 95th percentile (P95). 
    Here the $\Delta t$ step is 15 minutes.}
    \label{fig:VestaNoErr2}
\end{figure}

We then allow $\Delta t$ to take values in the interval $I_2$, which corresponds to wider arcs of observation. 
For each $\Delta t$ selected in $I_2$ we compute the errors in the angular momentum vector and its direction. 
Statistical quantities related to these errors are shown in Figure \ref{fig:VestaNoErr2} as functions of $\Delta t$. 
M$_{top}$ continues to show better results and a smoother behavior for values of $\Delta t$ smaller than 30 days, even if the improvement over M$_{geo}$ is less pronounced with respect to that shown in Figure \ref{fig:VestaNoErr1}. 
In such interval, the geocentric method is much more sensitive to $\Delta t$ and large oscillations having a period of one day appear. 
This is due to not accounting for the topocentric position of the observer. 
We also notice that both methods exhibit an optimal performance for $\Delta t\approx$ 3 weeks, and M$_{top}$ obtains in $75\%$ of the solutions an error smaller than $0.2\%$ and $0.03\%$ in the vectors ${\bf c}$ and $\hat{{\bf c}}$, respectively.

If we consider time intervals longer than 30 days, the performance of the two methods is almost comparable, which is expected because the terms introduced in the topocentric version become smaller as $\Delta t$ grows. 
The solutions with both methods deteriorate for $\Delta t > 80$~days since they rely on Taylor's expansions with respect to $\Delta t$.

\begin{table}[h!]
    \small
    \begin{center}
    \begin{tabular}{c|r|r|r|r|r|r}
        ${\Delta t}$ (days) &0.02 &1 &10 &50 &100 & True value\\
        \hline
        $i_1$ (geo) &17.80160   &7.06410   &7.20588   &7.09387   &7.05670   &\multirow{4}{*}{7.14165}\\
        $i_2$ (geo) &0.00315    &0.00297   &0.00303   &0.00436   &0.00278   &\\
        $i_1$ (top) &7.14611    &7.06413   &7.20572   &7.09407   &7.05700   &\\
        $i_2$ (top) &17.79875   &0.00156   &0.00328   &0.00305   &0.00881   &\\
        \hline
        $\Omega_1$ (geo) &150.28159   &103.18144   &104.36892      &103.20691   &   105.07597 & \multirow{4}{*}{103.80838}\\
        $\Omega_2$ (geo) &140.61411   &146.13681   &$-$143.70328   &160.84846   &   179.83017 & \\
        $\Omega_1$ (top) &103.83824   &103.17391   &104.35977      &103.19931   &   105.07223 & \\
        $\Omega_2$ (top) &150.27855   &167.05122   &$-$112.30086   &$-$174.23000   &$-$66.15466 & \\
        \hline
        $c_1$ (geo) &1125.88148   &0.02600   &0.02650   &0.02694   &0.03463   & \multirow{4}{*}{0.02633}\\
        $c_2$ (geo) & 0.01721     &0.01721   &0.01719   &0.01720   &0.01719   & \\
        $c_1$ (top) & 0.02628     &0.02600   &0.02650   &0.02694   &0.03463   & \\
        $c_2$ (top) &1124.55534   &0.01720   &0.01719   &0.01720   &0.01718   & \\
        \hline
    \end{tabular}
    \end{center}
    \caption{Values of the inclination ($i$), longitude of the node ($\Omega$), magnitude of the angular momentum vector ($c$), obtained by M$_{top}$ and M$_{geo}$, with observations of Vesta not affected by astrometric error and with different time intervals between consecutive observations. 
    The epoch of the true orbit is 59200 MJD, the epoch of the first observation is $\sim$ 59200.012 MJD (due to aberration correction), and the angles are in degrees.}
    \label{tab:geo_noerr}
\end{table}
\normalsize

In Table~\ref{tab:geo_noerr} we report the inclination, longitude of the node, and magnitude of the angular momentum vector obtained by M$_{geo}$ and M$_{top}$ from observations of Vesta without astrometric error, considering different values of $\Delta t$ with the same epoch for the first observation. 
We also show for M$_{geo}$ with the label ``2'' the solution of Mossotti's original method which is always discarded, i.e. the one for which the angular momenta of the asteroid and the Earth are equal.
We observe that if $\Delta t$ is large enough, the values of $i$ and $c$ of this spurious solution are close to the \emph{wrong} solution from M$_{top}$ (also labeled by ``2''). The same behavior is not observed for $\Omega$ because the Earth orbital inclination is small and therefore small variations in $i$ can cause large deviations in $\Omega$.
For $\Delta t = 30$ minutes ($0.02$ days) the solution from M$_{geo}$ with label ``1'' is very close to the wrong solution from M$_{top}$: in this case only M$_{top}$ gives values of $i$, $\Omega$, $c$ close to the true ones.

\subsection{Tests with astrometric errors}
\label{s:error}

\begin{figure}[ht!]
    \centering
    \includegraphics[trim= 0mm 0mm 0mm 0mm, clip, width=0.97\textwidth]{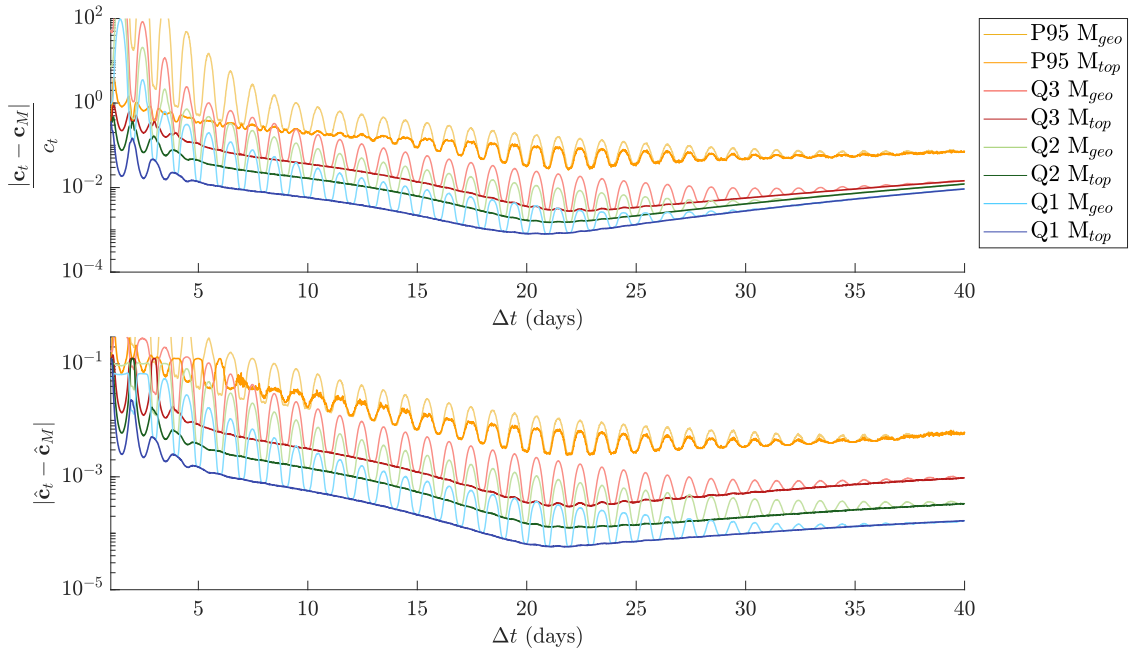}
    \caption{Same as Figure \ref{fig:VestaNoErr2} but the observations of Vesta are affected by astrometric error with zero mean and rms of $0.1$ arcsec. 
    Here the $\Delta t$ step is 15 minutes.}
    \label{fig:Vesta1e-1Err2}
\end{figure}

The same numerical tests described in the previous section for the asteroid Vesta are carried out by introducing an astrometric error with zero mean and standard deviation (rms) of 0.1 arcsec in the values of right ascension and declination, which is typical of modern asteroid surveys like Pan-STARRS1.
Neither M$_{geo}$ nor M$_{top}$ gives reliable results for $\Delta t$ between 15 and 200 minutes.
Indeed, determining a preliminary orbit from a single short arc is a challenging task, sometimes impossible without considering infinitely many solutions \cite{multsol}, \cite{admisreg}. 
However, the computation of a preliminary orbit can be performed by linking together two or more short arcs (e.g. \cite{gbm15}, \cite{gbm17}). 
On the other hand, both methods yield satisfactory results for time intervals between two consecutive observations larger than 8 hours with only a slight degradation of their performance with the introduction of astrometric error (compare Figures~\ref{fig:VestaNoErr2} and \ref{fig:Vesta1e-1Err2}). 
As in the case without astrometric error, M$_{top}$ is better than M$_{geo}$ for any considered $\Delta t$. 
A smoother behavior of M$_{top}$ is observed for $\Delta t$ smaller than 30 days. 
Both methods perform best for $\Delta t \approx$ 3 weeks.

\begin{figure}[ht]
    \centering
    \includegraphics[trim= 0mm -5mm 0mm 0mm, clip, width=0.45\textwidth]{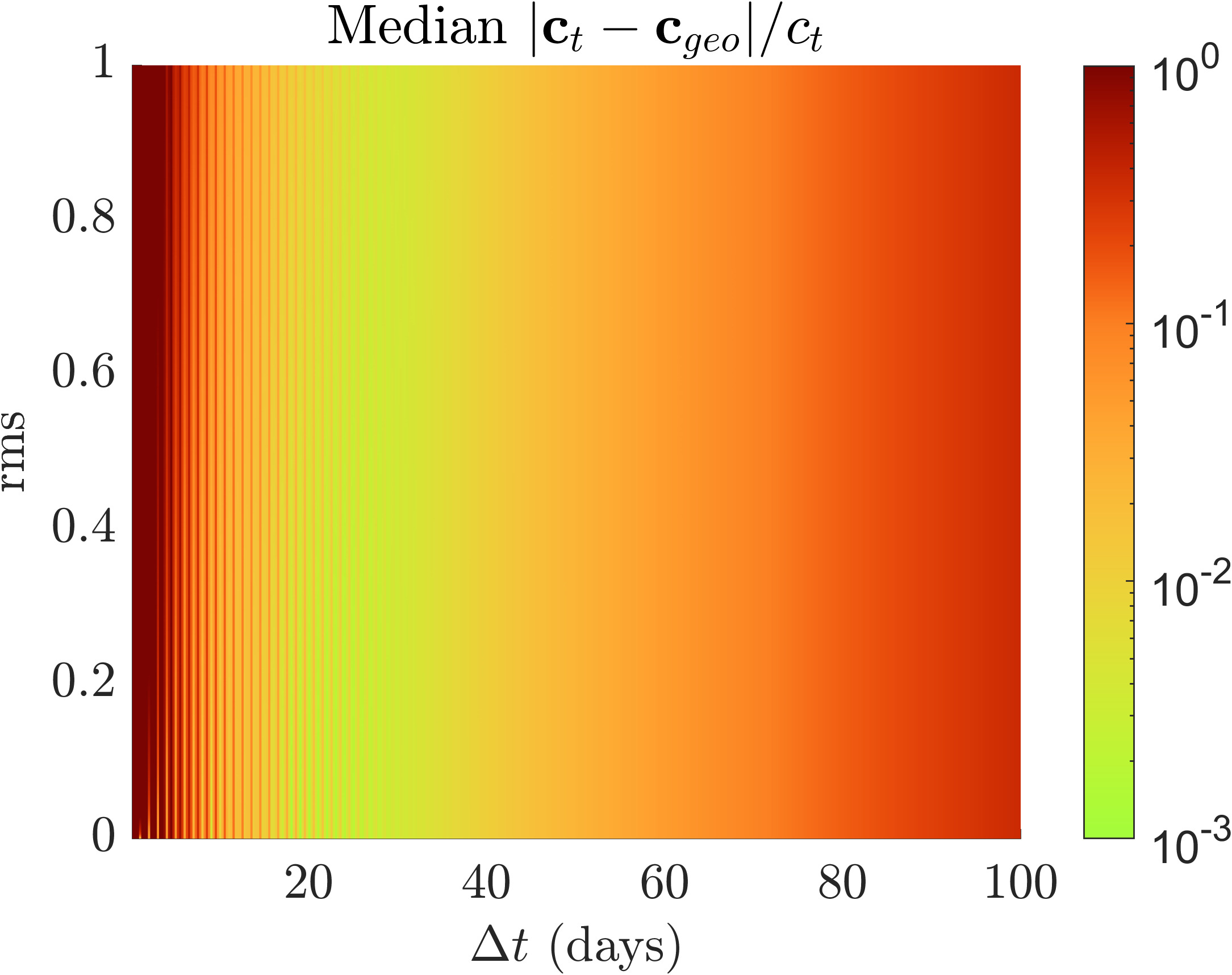}
    \includegraphics[trim= 0mm -5mm 0mm 0mm, clip, width=0.45\textwidth]{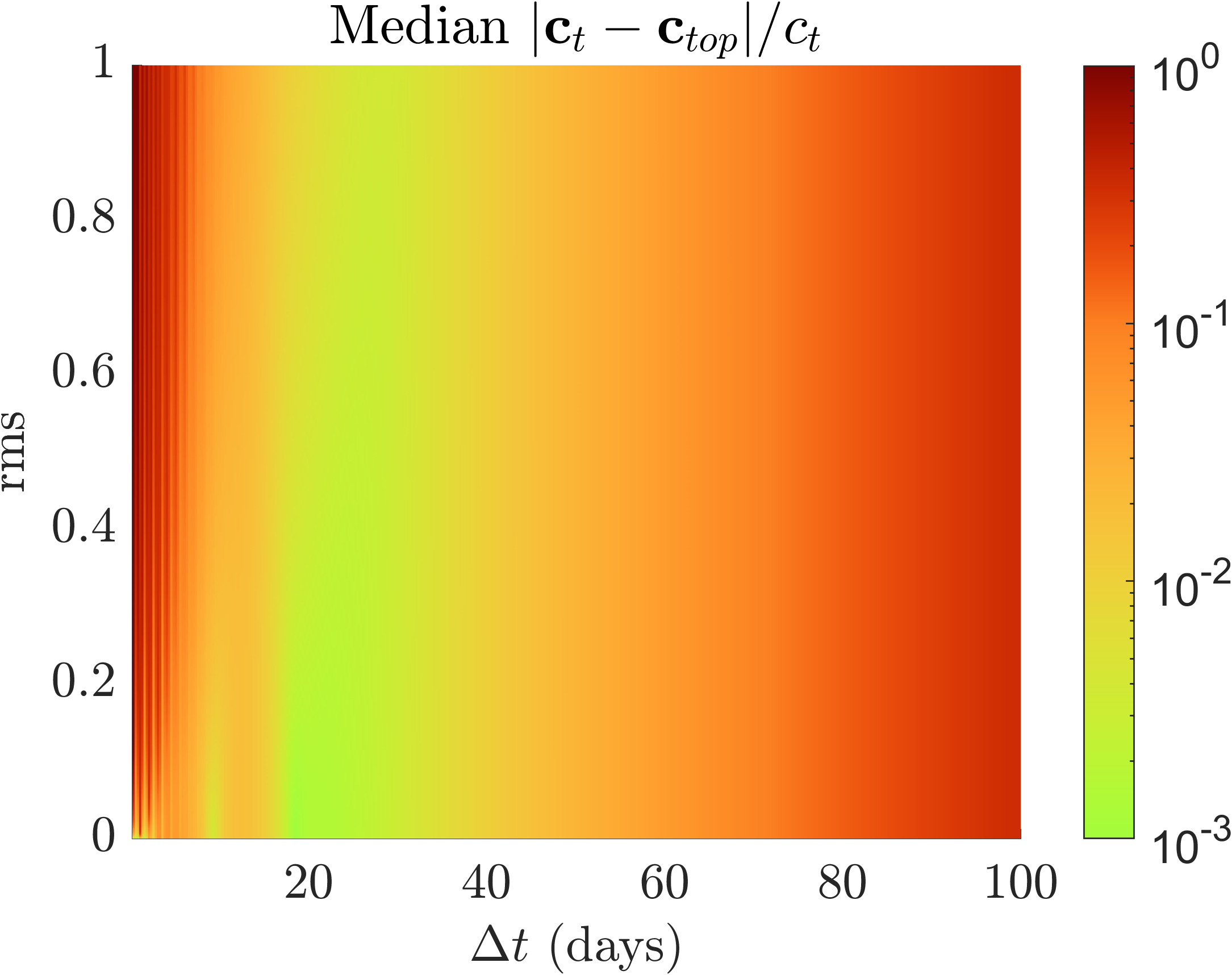}
    \includegraphics[trim= 0mm -5mm 0mm 0mm, clip, width=0.45\textwidth]{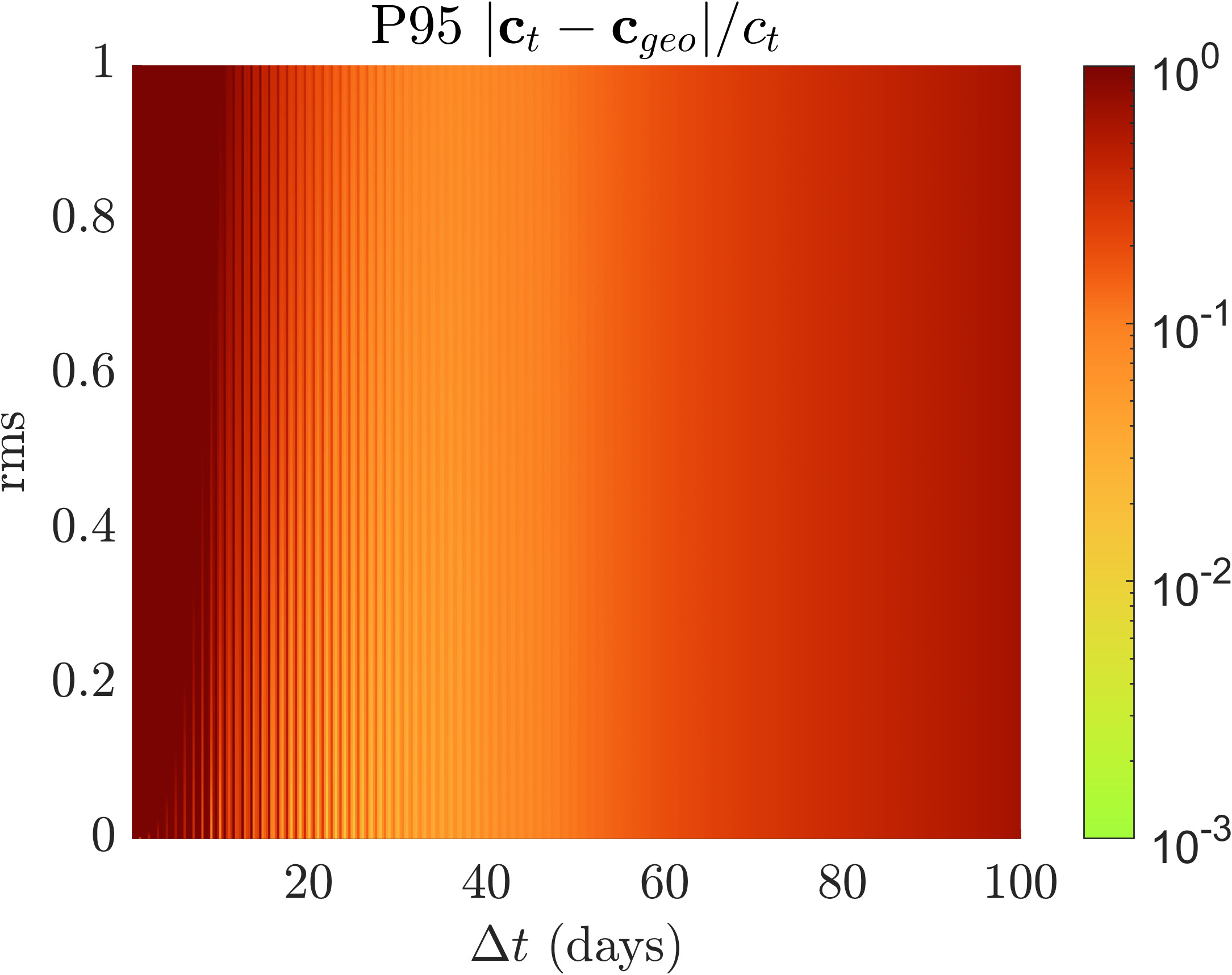}
    \includegraphics[trim= 0mm -5mm 0mm 0mm, clip, width=0.45\textwidth]{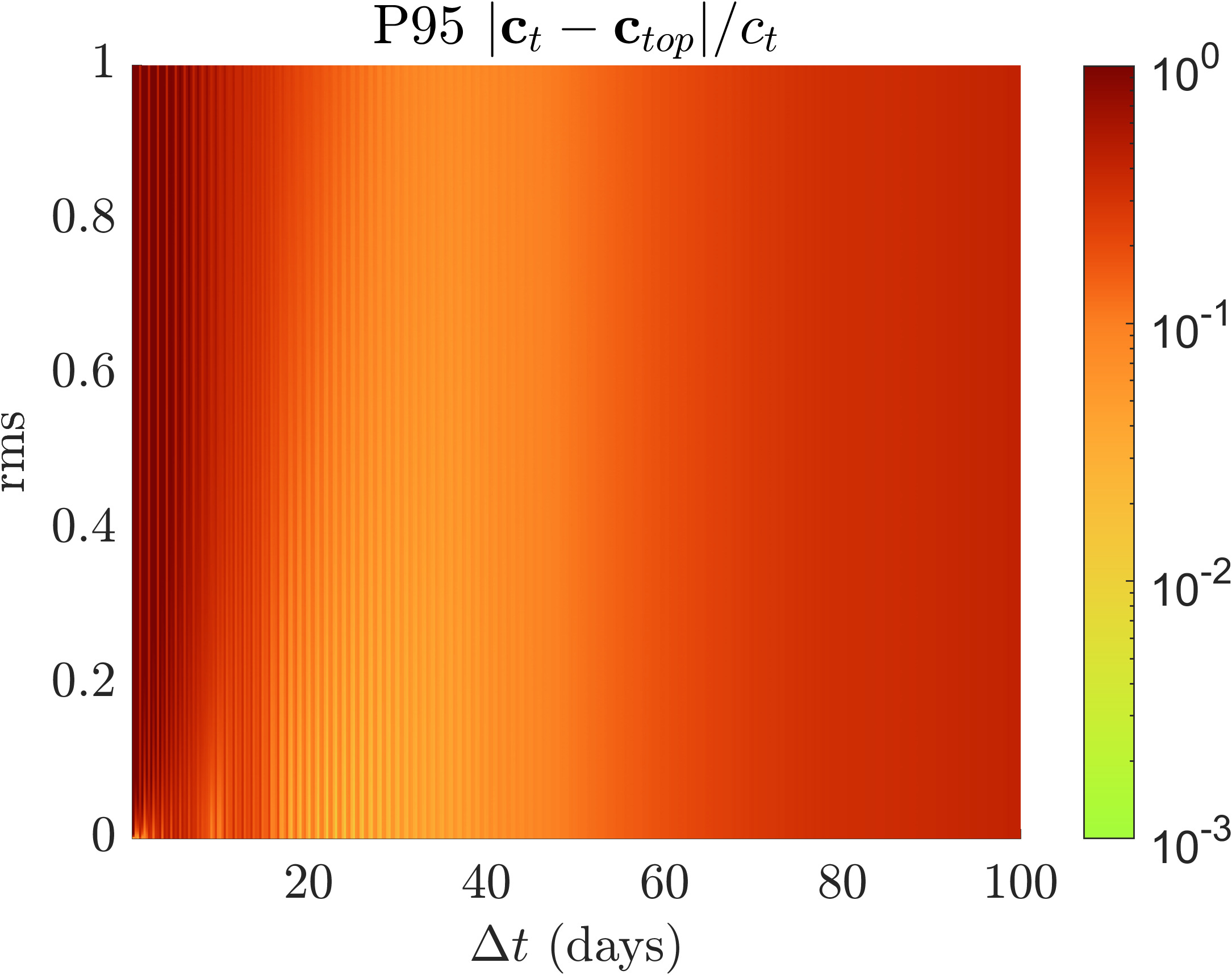}
    \vspace{-3mm}
    \caption{Performance of Mossotti's original method (\emph{left}) and its topocentric version (\emph{right}) with observations of 546077 numbered asteroids. 
    For each asteroid we compute the difference between the true angular momentum vector ${\bf c}_t$ at the epoch 59200 MJD, and the same quantity ${\bf c}$ obtained by either M$_{geo}$ or M$_{top}$ on a uniform grid of values of $\Delta t$ and rms of the astrometric error in the intervals $[0.25,\,100]$ days and $[0,\,1]$ arcsec. 
    Colored representations of the median and the 95th percentile related to the error $|{\bf c}_t-{\bf c}|/c_t$ are displayed.
    The step of $\Delta t$ is 6 hours, the step of rms is $0.002$ arcsec.}
    \label{fig:cvecdays}
\end{figure}

\begin{figure}[ht!]
    \centering
    \includegraphics[trim= 0mm -5mm 0mm 0mm, clip, width=0.45\textwidth]{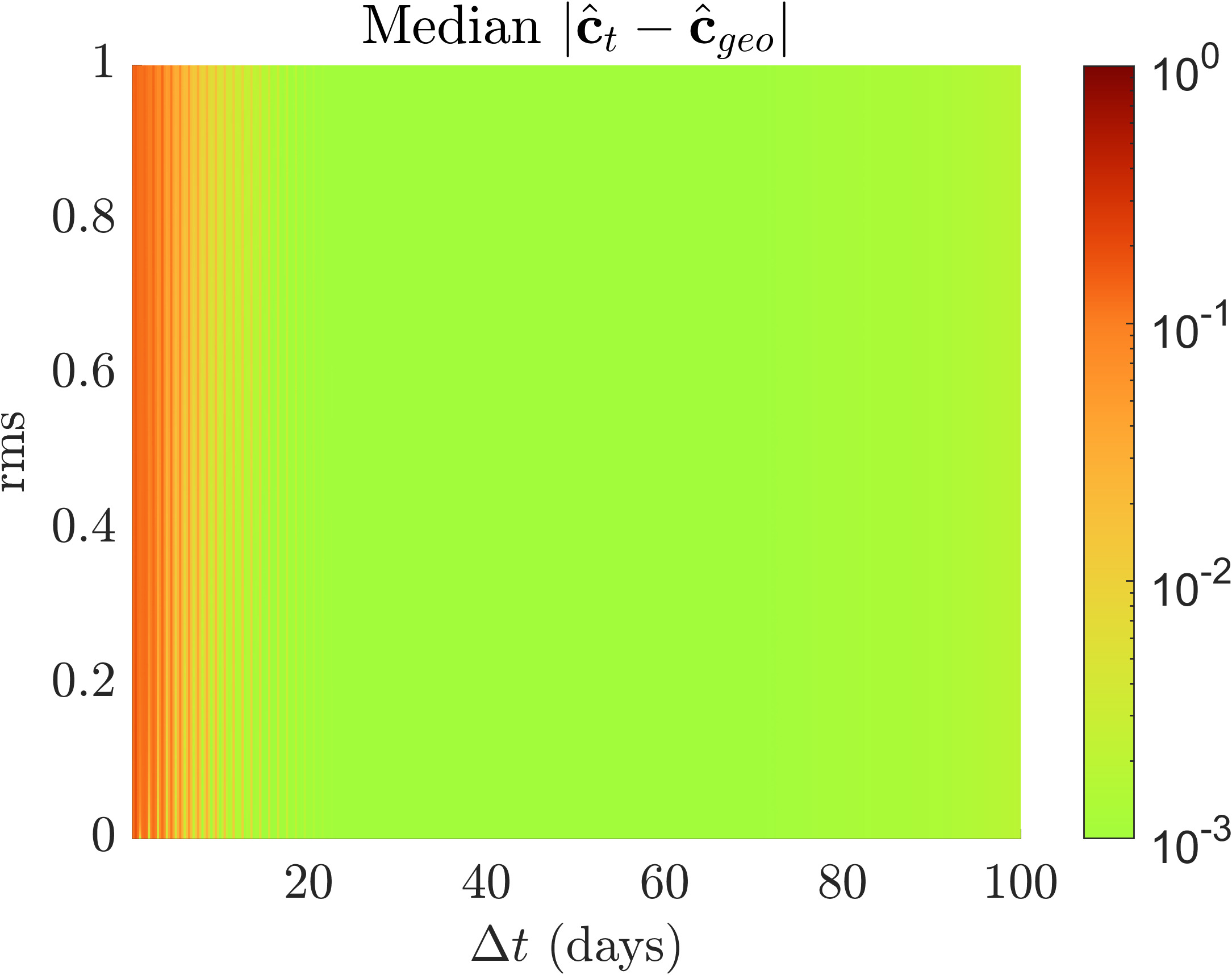}
    \includegraphics[trim= 0mm -5mm 0mm 0mm, clip, width=0.45\textwidth]{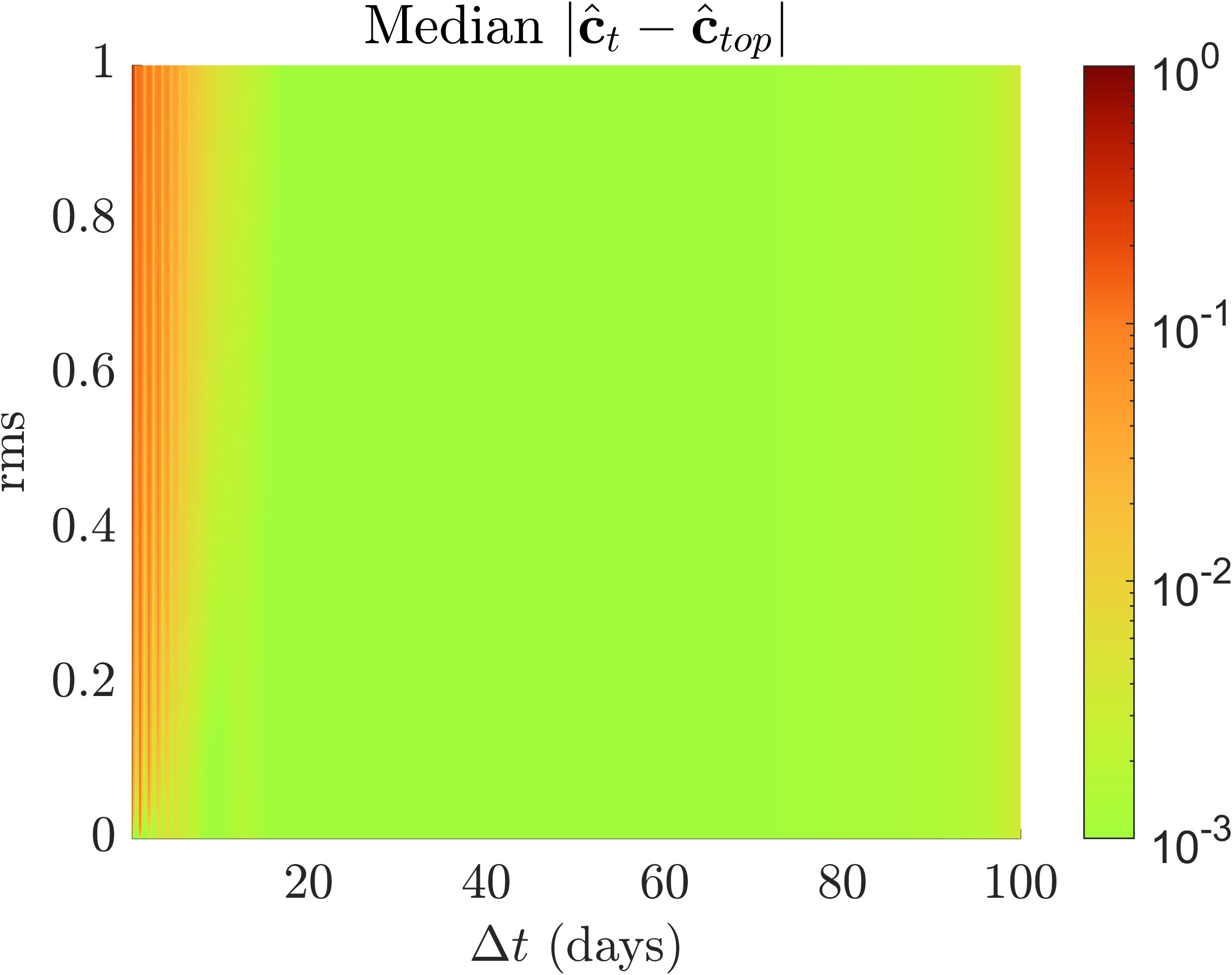}
    \includegraphics[trim= 0mm -5mm 0mm 0mm, clip, width=0.45\textwidth]{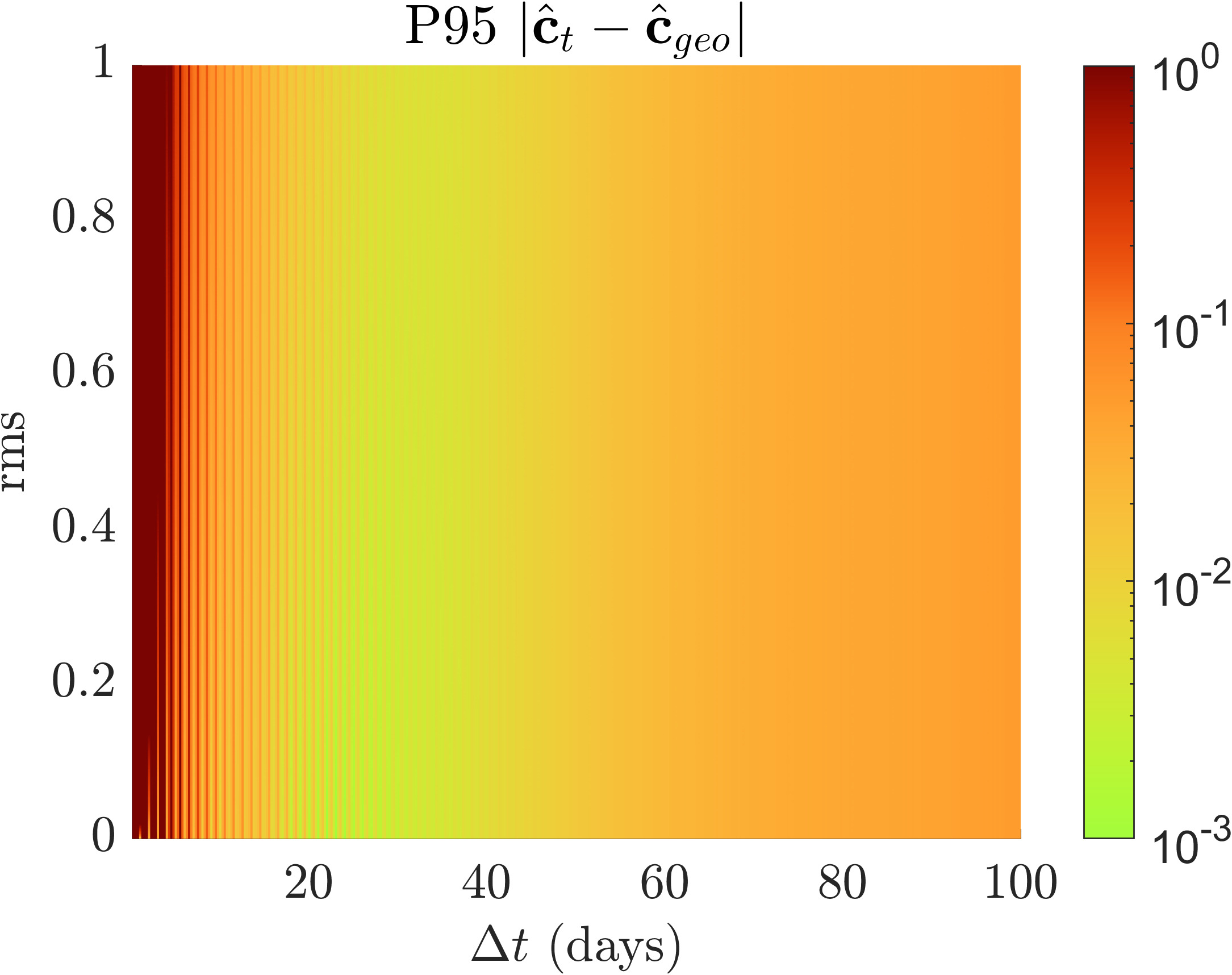}
    \includegraphics[trim= 0mm -5mm 0mm 0mm, clip, width=0.45\textwidth]{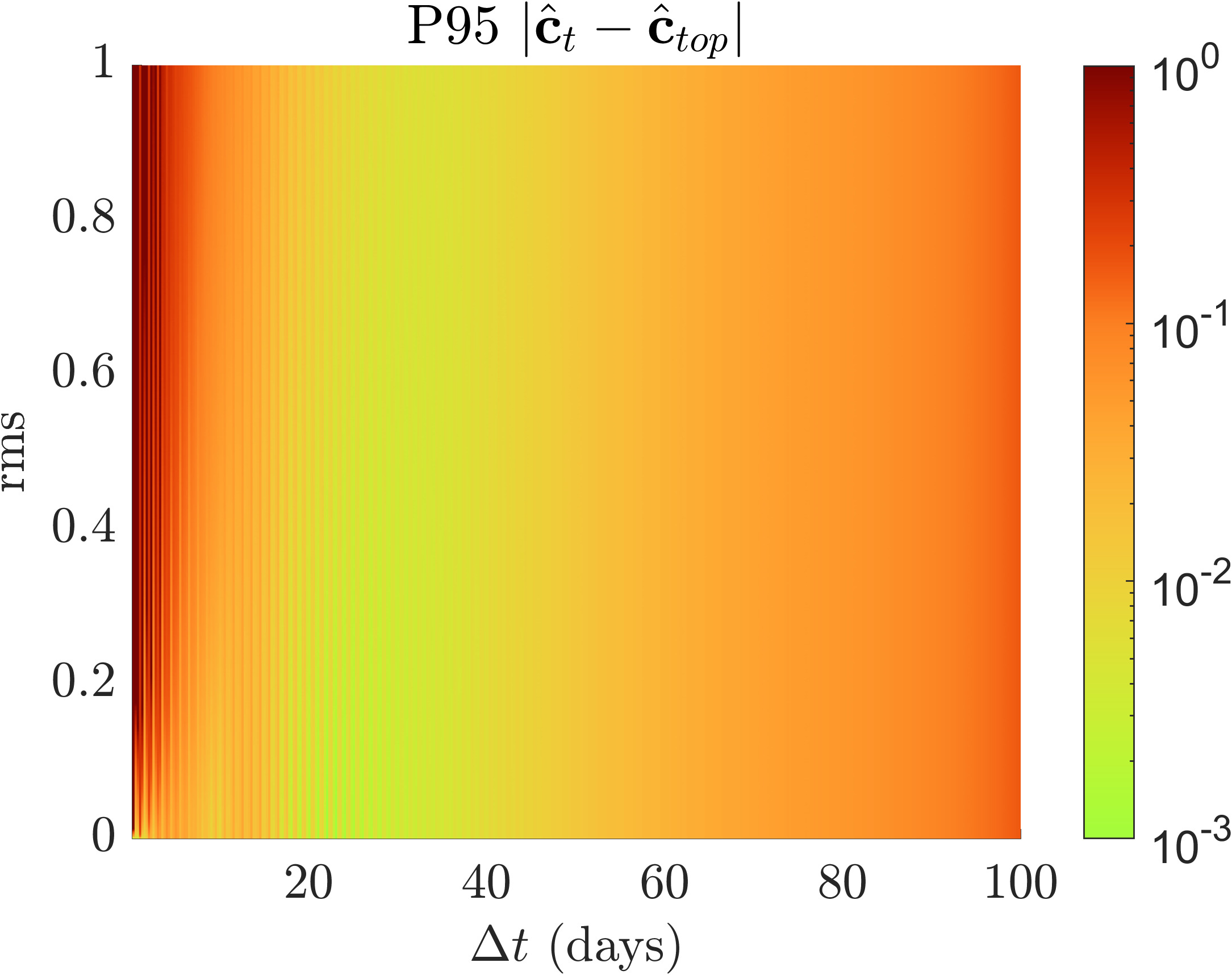}
    \vspace{-3mm}
    \caption{Same as Figure~\ref{fig:cvecdays} for the error in the direction of the angular momentum vector.}
    \label{fig:cdirdays}
\end{figure}

In our third test we use M$_{geo}$ and M$_{top}$ with synthetic observations of the 546077 numbered asteroids known to the date of October 19, 2020. 
Their orbital elements at the epoch 59200 MJD (from AstDyS-2) provide the true orbit and are used to simulate observations from the Pan-STARRS1 telescope site.
The epoch of the first observation is obtained from 59200 MJD considering aberration correction; then the subsequent three observations are simulated by Keplerian propagation. 
The time $\Delta t$ is varied in the interval $I_2$ given in \eqref{I12}, and the standard deviation of the astrometric error spans from 0 to 1 arcsec.
Figures~\ref{fig:cvecdays}, \ref{fig:cdirdays} show that for time intervals between 20 and 40 days both M$_{geo}$ and M$_{top}$ give good results. 
A closer look at this range of values of $\Delta t$ for M$_{geo}$ reveal the same oscillations displayed in Figures \ref{fig:VestaNoErr2} and \ref{fig:Vesta1e-1Err2}. 
On the other hand, M$_{top}$ smooths out such oscillations. 
The best performance of both M$_{geo}$ and M$_{top}$ is reached again for $\Delta t\approx$ 3 weeks.

\subsection{Tests on synthetic survey data without astrometric errors}
\label{s:synthetic-survey-tests}

Our final test introduces more realism into the simulations in an attempt to assess the performance of the topocentric version of Mossotti's method (M$_{top}$) on synthetic data that uses a realistic cadence and accounts for actual observability of the asteroids. 
With the motivation that Mossotti's initial orbit determination method could be applied to linking observations of unknown asteroids in contemporary and future asteroid surveys, we applied M$_{top}$ to synthetic observations from the Vera Rubin Observatory's (VRO) Legacy Survey of Space and Time (LSST) \cite{Ivezic2008-LSST}. 
They have developed a high-fidelity survey scheduler that will be employed in final operations but is currently being used to simulate and optimize the survey strategy \cite{Connolly2014-LSSTsims,Delgado2016-LSSTscheduler,Naghib2018-LSSTscheduler}.
We used a single survey simulation for one month of surveying that did not include any astrometric error. 
Then we extracted the first four synthetic detections of all the detected numbered NEOs, Trojans, Centaurs, and TNOs, but only a small subset of the detected main belt objects so that they would not dominate our results. 
The epochs of observation were all within about 30 days and the astrometric positions were generated with a full $n$-body integration. 
This process produced a set of four detections of 1535 objects distributed throughout the solar system.  
We then processed all the detections with both M$_{top}$ and our implementation of Gauss's method, by limiting our search to bounded orbits only.

In our sample Gauss's method was able to produce orbits for 1493 objects (i.e. $\sim97$\%) and M$_{top}$ provided solutions for 1395 objects (i.e. $\sim91$\%). 
However, we had 59 occurrences of a negative discriminant of equation \eqref{eq:sctop} with M$_{top}$, and for some of them we were able to recover an acceptable orbit by setting the discriminant equal to zero.  
Doing so increases the number of solutions for M$_{top}$ to 1454 (i.e. $\sim95$\%).
  
While Gauss's method can yield three different solutions for the same set of observations and M$_{top}$ can yield two, on this set of data they had multiple solutions for about 50\% and 45\% of the objects, respectively. 
Gauss's method did not produce any orbit for 42 objects and M$_{top}$ for 81. 
Moreover, both of them failed in 30 of these cases. 
In 12 cases M$_{top}$ was able to obtain at least one orbit while Gauss's method was not, and for some cases it found an acceptable orbit.

The primary benefit of M$_{top}$ is that in our limited testing it appears to be about $6$ times faster than Gauss's method. 
We repeated the orbit computation 1000 times for each of the 1535 objects using an Intel Xeon processor, with base clock 3.30 GHz: Gauss's algorithm took $\sim 77$ seconds, while M$_{top}$ $\sim 13$ seconds.
 
\begin{figure}[ht!]
    \centering
    \includegraphics[width=0.32\textwidth]{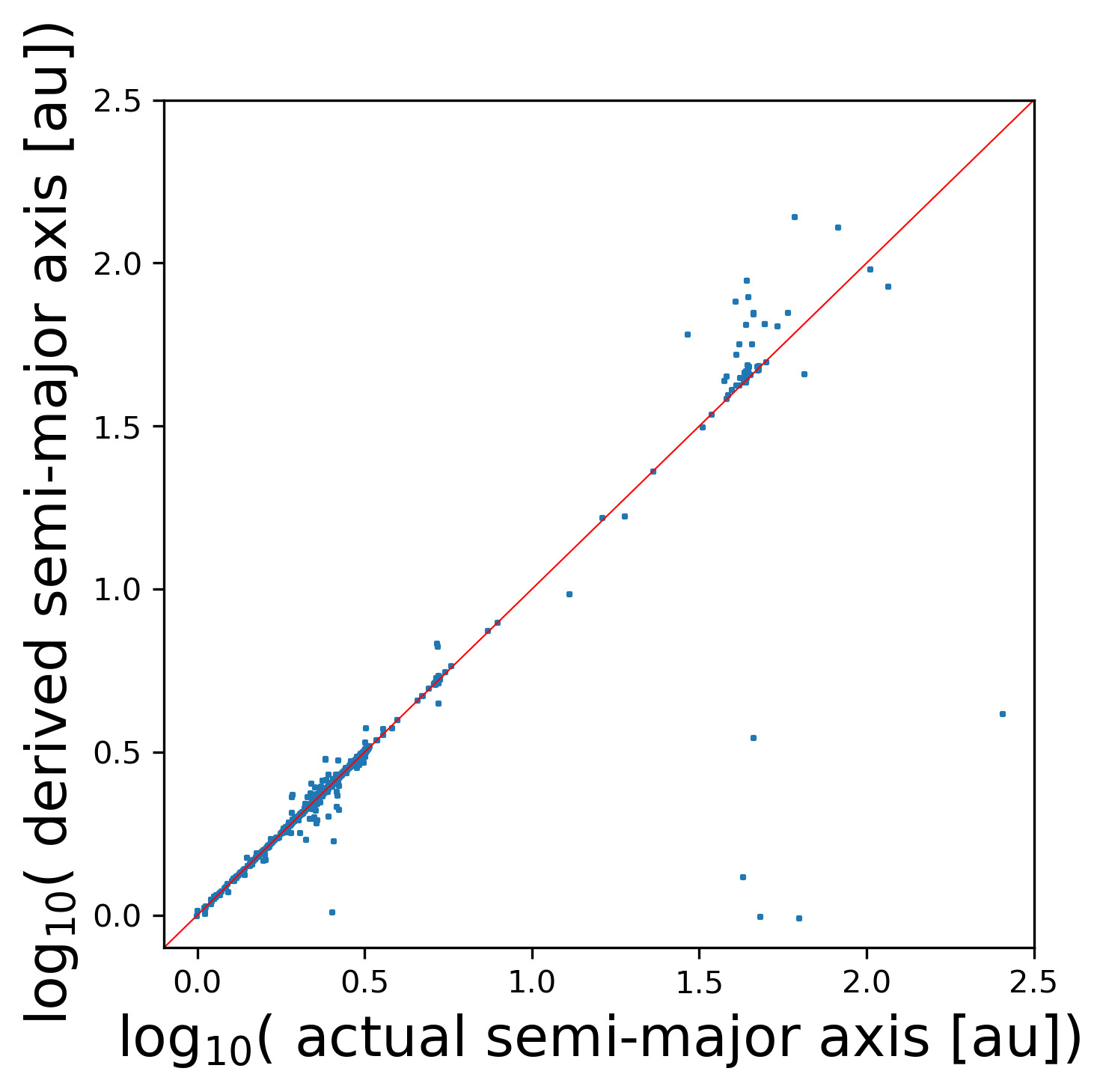}
    \includegraphics[width=0.32\textwidth]{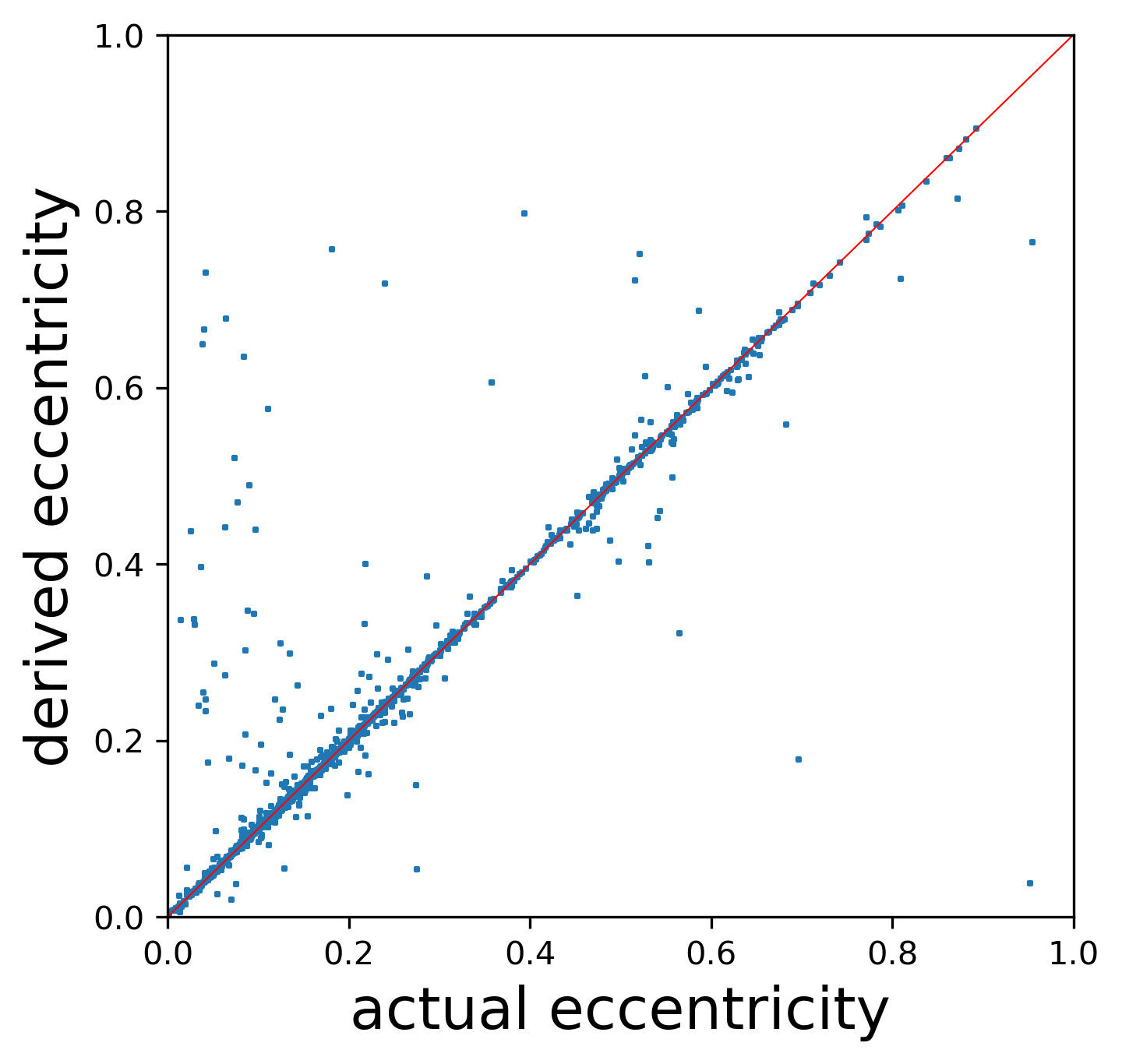}
    \includegraphics[width=0.32\textwidth]{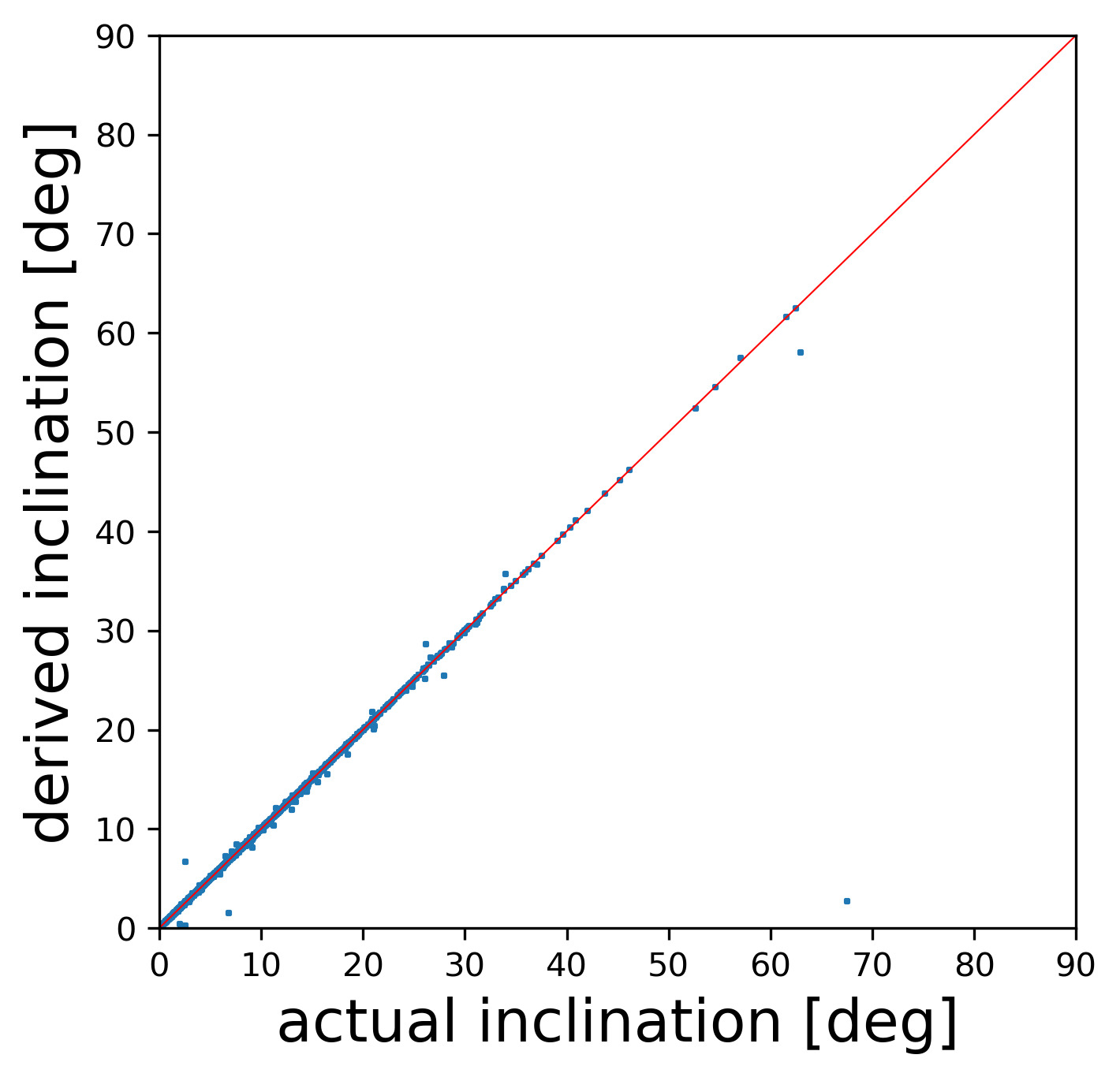}
    \includegraphics[width=0.32\textwidth]{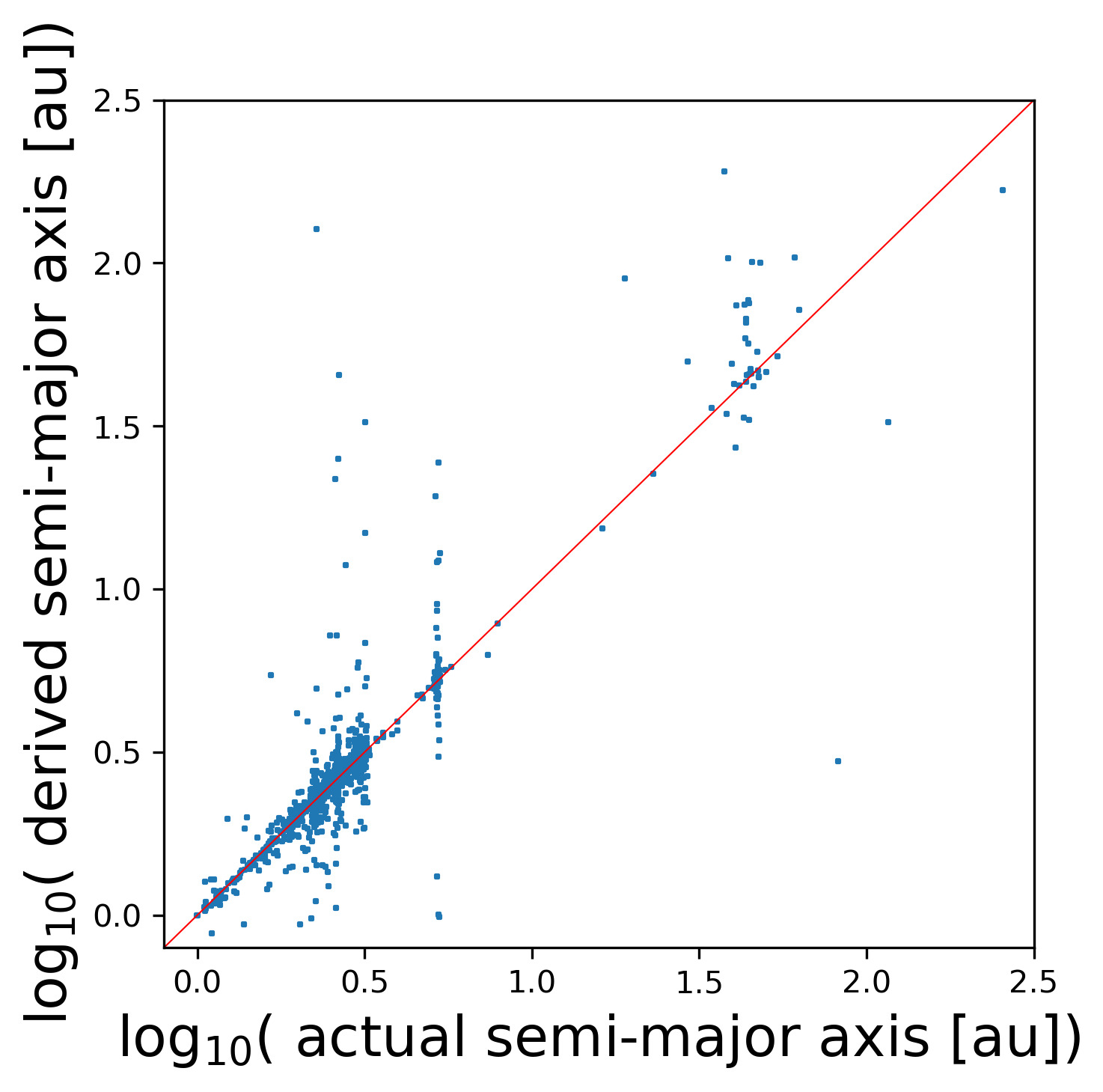}
    \includegraphics[width=0.32\textwidth]{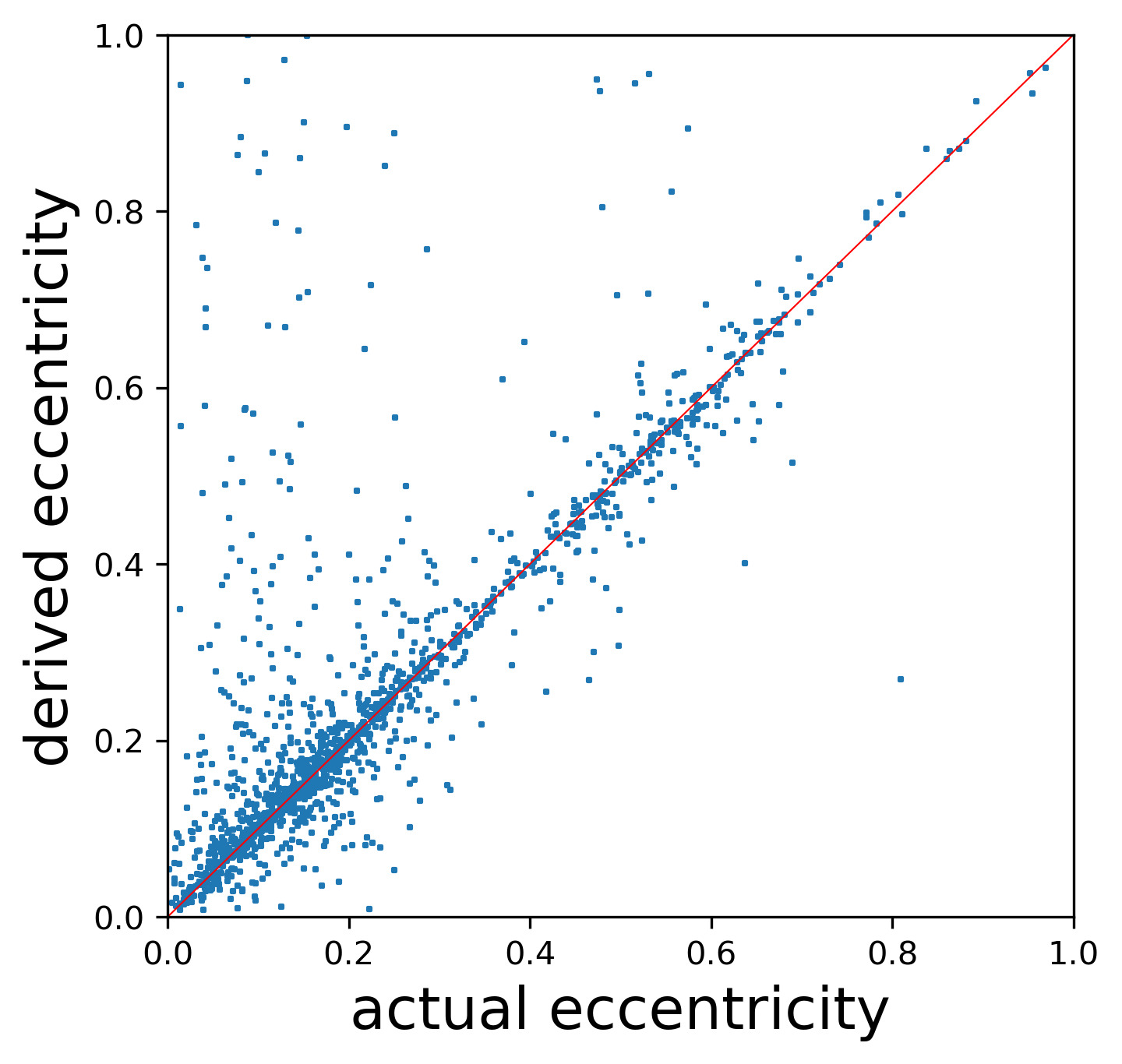}
    \includegraphics[width=0.32\textwidth]{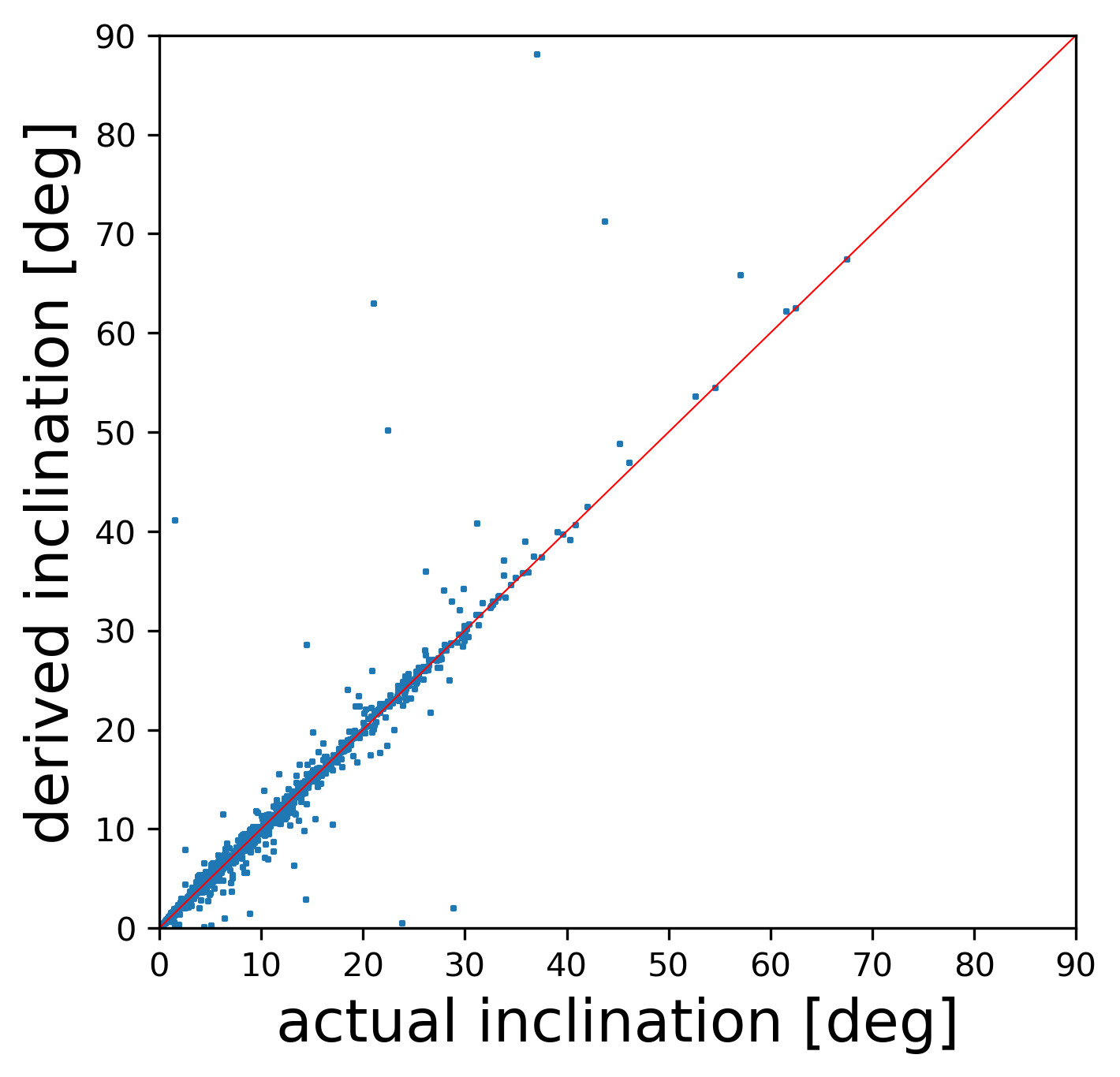}
    \caption{All the panels present the derived orbital parameter on the $y$-axis versus the actual parameter value on the $x$-axis using error-free data from a VRO-LSST simulation. 
    The top panels are for Gauss's method and the bottom panels use the topocentric version of Mossotti's method. 
    From left to right the panels compare $\log_{10}(a/[au])$, where $a$ is the semi-major axis, eccentricity, and inclination. The red reference line in each figure has a slope of 1.}
    \label{fig:aei-vs-aei}
\end{figure}

Gauss's technique provides better solutions for objects throughout the solar system (Figure \ref{fig:aei-vs-aei} and Table \ref{tab:gauss-mossotti-comparison}).  
When comparing the derived orbital elements with their actual values we used the derived orbit solution that had the lowest 5-element $D$-criterion\footnote{The $D$-criterion quantifies the difference between two orbits using all the orbital elements except for the mean anomaly.} relative to the actual orbit \cite{Drummond1981-Dcriterion}.
A simple visual comparison of the results suggests that Gauss's method is more likely to produce good orbital elements and less likely to yield wildly different values. 
Quantitative orbital element comparisons confirm this impression (Table \ref{tab:gauss-mossotti-comparison}).

\begin{table}[ht!]
    \centering
    \begin{tabular}{|l|r|r|r|r|}
        \hline
        & \multicolumn{2}{|c|}{Gauss}  & \multicolumn{2}{|c|}{Mossotti} \\
        \hline
        orbital element        & $\bar\Delta$  &   rms  & $\bar\Delta$ & rms  \\
        \hline
        semi-major axis (au)   &    0.13  &   3.44  &   0.28  &   3.18  \\
        eccentricity           &    0.01  &   0.07  &   0.03  &   0.12  \\
        inclination (deg)      & $-$0.02  &   0.27  &   0.05  &   2.31  \\
        long. node (deg)       & $-$0.09  &   1.83  &   0.10  &   4.58  \\
        arg. perihelion (deg)  & $-$0.38  &   8.51  &   0.10  &  19.17  \\
        \hline
    \end{tabular}
    \caption{Mean and rms of the difference between the actual and  derived orbital elements for both Gauss's method and the topocentric version of Mossotti's method. 
    For the angular elements we eliminated some outliers to better illustrate the difference between the methods in the vast majority of cases.}
    \label{tab:gauss-mossotti-comparison}
\end{table}

\section{Conclusions}
In this paper we have revisited Mossotti's orbit determination method working with four geocentric observations of a celestial body, and extended it to the case of topocentric observations. 
While Mossotti's method yields linear equations for the components of the angular momentum vector, the topocentric version leads to a quadratic
equation. 
Numerical simulations with synthetic observations both without and with astrometric error show that the topocentric method improves the original one. 
Considering all the numbered asteroids, and generating for each of them four observations equally spaced in time, we find that both these methods show an optimal behavior for a time separation $\Delta t$ between two consecutive observations of about 3 weeks. 
Finally, we compare the new method with Gauss's method using synthetic observations without astrometrical error that reproduce the expected scheduling of the Vera Rubin Observatory's (VRO) Legacy Survey of Space and Time (LSST), characterized by an average $\Delta t$ of about 4 days. 
Gauss's method provides good orbits for a larger number of objects than the topocentric version of Mossotti's method, which, on the other hand, is faster.

\section{Acknowledgments}
We thank Dr. Lynne Jones, Dr. Siegfried Eggl, Dr. Sam Cornwall, and Dr. Mario Juri\'c of the University of Washington (WA) for assistance in identifying, accessing, and understanding the appropriate VRO/LSST simulations. 
We also thank the anonymous referees for their useful comments.
GFG and GB acknowledge the project MIUR-PRIN 20178CJA2B titled ``New frontiers of Celestial Mechanics: theory and applications".  
GFG, GB and OR have been partially supported by the MSCA-ITN Stardust-R, Grant Agreement n. 813644 under the H2020 research and innovation program.

\appendix
\section{Comparison with Mossotti's original paper}
\label{append}

\setlength\extrarowheight{2pt}
\begin{table}[ht!]
\centering
\begin{tabular}{>{\centering\arraybackslash}p{0.185\textwidth}>{\centering\arraybackslash}p{0.15\textwidth}>{\centering\arraybackslash}p{0.04\textwidth}>{\centering\arraybackslash}p{0.24\textwidth}>{\centering\arraybackslash}p{0.21\textwidth}}
\cline{1-2}\cline{4-5}
\textbf{Mossotti} & \textbf{here} &  & \textbf{Mossotti} & \textbf{here}\\
\cline{1-2}\cline{4-5}
$(x,\,y,\,z)^t$ & $\bm{r}$ &  
& $(Q''',\,-Q'',\,Q')^t$
& $\displaystyle\frac{1}{\theta_{12}^{2}}\Bigl(\frac{u_1}{\theta_{23}}, \frac{u_2}{\theta_{31}}, \frac{u_3}{\theta_{12}}\Bigr)^t$ \\
$(X,\,Y,\,Z)^t$ & $\bm{q}$ & 
& \multirow{3}{=}{$\left(\begin{matrix}B_1''' & B_2''' & B_3'''\\ -B_1'' & -B_2'' & -B_3''\\ B_1' & B_2' & B_3'\\ \end{matrix}\right)$} &\\ 
$(\mu,\,\nu,\,\omega)^t$ & $\hat{\bm{q}}$ & 
& & $\adj(\hat{Q})\hat{P}$ \\ 
$R,\,D$ & $q$ & 
& &\\ 
$(m,\,n,\,o)^t$ & $\hat{\bm{\rho}}$ & 
&  \multirow{3}{=}{$\hspace{2.5mm}\left(\begin{matrix}b_1''' & b_2''' & b_3'''\\ -b_1'' & -b_2'' & -b_3''\\ b_1' & b_2' & b_3'\\ \end{matrix}\right)$} &\\ 
$\delta$ & $\rho$ &
& & $\adj(\hat{P})\hat{Q}$\\ 
$\kappa(C''',\,-C'',\,C')^t$ & $\bm{c}_\oplus$ & 
& &\\ 
$\kappa(c''',\,-c'',\,c')^t$ & $\bm{c}$ & 
& $(\psi',\,\psi'',\,\psi''')$&$\frac{1}{\kappa}\bm{c}^t\hat{P}$\\ 
$(T''',\,-T'',\,T')^t$ & $\bm{T}$ & 
& $(\chi',\,\chi'',\,\chi''')$&$\frac{1}{\kappa}\bm{c}_\oplus^t\hat{P}$\\ 
$(\tau''',\,-\tau'',\,\tau')^t$ & $\bm{\tau}$ & 
& $(\varphi',\,\varphi'',\,\varphi''')$&$\frac{1}{\kappa}(\bm{c}_\oplus-\bm{c})^t\hat{Q}$\\ 
$(\theta''',\,-\theta'',\,\theta')^t$ & $\bm{\theta}$ &
& $(\Phi',\,\Phi'',\,\Phi''')$&$\frac{1}{\kappa}(\bm{c}_\oplus-\bm{c})^t\hat{P}$\\ \cline{1-2}\cline{4-5}\\
    \end{tabular}
    \caption{Correspondence table between Mossotti's and our notation.}
  \label{compare}
\end{table}

\bibliography{mybib}{}
\bibliographystyle{plain}

\end{document}